\documentclass[a4paper,11pt]{article}
\pdfoutput=1

\usepackage[english]{babel}

\usepackage{float}
\usepackage{jheppub}
\usepackage{amsmath,amsfonts,amssymb,amsthm,graphicx,color}
\usepackage{mathrsfs}
\usepackage{enumerate}
\usepackage{epsfig}
\usepackage{multirow}
\usepackage{comment}
\usepackage{datetime}
\usepackage{slashed}
\usepackage{framed}
\usepackage{longtable}
\usepackage[mathscr]{euscript}
\usepackage{cancel}
\usepackage{tensor}
\usepackage{mathtools}
\usepackage{caption}
\usepackage{subcaption}
\usepackage[multiple]{footmisc}
\usepackage[T1]{fontenc}
\usepackage[utf8]{inputenc}
\usepackage{placeins}
\usepackage{hyperref}
\usepackage{multirow}
\usepackage{graphicx}

\usepackage{bbm}
\usepackage{mathtools}
\usepackage{braket}
\usepackage{mathrsfs}
\usepackage{tensor}
\usepackage{cancel}
\usepackage{slashed}
\usepackage{booktabs}
\usepackage{physics}

\usepackage{enumitem}
\usepackage{verbatim}
\usepackage{tikz}

\def\N{\mathcal{N}}
\def\R{\mathbb{R}}

\title{\Large{Massive IIA flux compactifications with dynamical open strings}}

\author[a]{Juan Ram\'on Balaguer,}
\author[b,c]{Valentina Bevilacqua,}
\author[b,c]{Giuseppe Dibitetto,}
\author[d]{Jose J. Fern\'andez-Melgarejo,}
\author[b,c]{Giuseppe Sudano}
\date{}

\affiliation[a]{Departamento de F\'isica, Universidad de Murcia, Campus de Espinardo, E-30100 Murcia, Spain}
\affiliation[b]{Dipartimento di Fisica, Universit\`a di Roma ``Tor Vergata", Via della Ricerca Scientifica 1, 00133, Roma, Italy}
\affiliation[c]{INFN, Sezione di Roma2, Via della Ricerca Scientifica 1, 00133, Roma, Italy}
\affiliation[d]{Departamento de Electromagnetismo y Electr\'onica, Universidad de Murcia, Campus de Espinardo, E-30100 Murcia, Spain}

\emailAdd{juanramon.balaguer@um.es}
\emailAdd{valentina.bevilacqua@roma2.infn.it}
\emailAdd{dibitetto@roma2.infn.it}
\emailAdd{melgarejo@um.es}
\emailAdd{giuseppe.sudano@roma2.infn.it}

\abstract{
    We consider massive type IIA compactifications down to 4 dimensions in presence of O6 planes and D6 branes parallel to them, in order to preserve half-maximal supersymmetry in 4D. The dynamics of open strings living on the spacetime filling branes is taken into account, in the gauged supergravity description, by adding extra vector multiplets and embedding tensor components. The scalar potential gets new terms that can be matched with contributions coming from dimensional reduction of the non-Abelian DBI and WZ brane actions. In this setting, we analyze the vacuum structure of the theory and find novel AdS$_4$ vacua, both supersymmetric and non-supersymmetric ones.
    Furthermore, we address their perturbative stability by computing their mass spectra. Some of the vacua are found to be perturbatively stable, despite their being non-supersymmetric.
    We conclude by discussing the reliability of our setup in terms of higher-derivative corrections.
}

\begin{document}
\maketitle
\flushbottom 
\section{Introduction}
Embedding consistent low energy effective theories within string theory is one of the most challenging tasks in theoretical high energy physics. Generically, this involves dimensional reductions and possibly spontaneous supersymmetry breaking. The wide range of low energy effective models obtained through such constructions are by definition UV consistent and constitute the so-called string landscape. Some portions of the landscape are widely explored and contain various models with different amounts of supersymmetry in diverse spacetime dimensions. 

However, as opposed to the above \emph{top-down} approach, one could also  adopt a \emph{bottom-up} approach instead. This would consist in analyzing  different lower dimensional and possibly supersymmetric constructions and try to address the following question: \emph{do they satisfy all UV consistency requirements?} In case of a positive answer and only then, can these  be  coupled to quantum gravity in a UV regime. This is the typical approach advocated by the so-called Swampland Program \cite{Vafa:2005ui,Ooguri:2006in}, whose aim is to formulate a complete set of consistency conditions that any effective model must comply with, in order for its UV completion to exist at all.

If we only focus on theories with extended supersymmetry, the landscape of all possibilities is so heavily constrained, that it sometimes even allows for an exhaustive classification. This is certainly the case for models enjoying maximal supersymmetry, \emph{i.e.} $32$ supercharges. In ten spacetime dimensions, the only consistent maximal supergravities turn out to be type IIA and IIB supergravities, which exactly happen to be the low energy descriptions of the associated superstring theories. This is the prime realization of \emph{string universality}.

If we now move to theories with half-maximal supersymmetry and still remain in 10D, the string universality principle was proven in \cite{Adams:2010zy} by showing that the coupling of pure $\mathcal{N}=1$ supergravity to a gauge sector is constrained by anomaly cancellation to realize either $\mathrm{SO}(32)$ or $\mathrm{E}_8\times\mathrm{E}_8$ as gauge groups. These in turn coincide with the (only) gauge groups that are realized within heterotic or type I superstring theories.

In the last few decades great developments taught us many things concerning theories with $16$ supercharges and this has put us in the right position to test the string universality principle in lower dimensions as well. At present, we may consider it to be fully settled in dimensions eight and higher \cite{Cvetic:2020kuw,Cvetic:2021sjm,Bedroya:2021fbu}. Beyond this, recent developments indicate that it might be verified even in dimension seven and six, the latter both in the chiral $(2,0)$ case \cite{Taylor:2019ots}, and in the non-chiral $(1,1)$ \cite{Fraiman:2022aik} case.  For $D<6$ cases, we do not have any conclusive arguments, but the preliminary indications look quite promising \cite{Fraiman:2018ebo,Font:2020rsk,Font:2021uyw}.

Our work is inspired by this whole context, from which it gets its chief motivation. Our specific goal is that of exploring the landscape of 4D string vacua with a dynamical contribution from the open string sector associated with the D-branes present in our setup. This work makes crucial use of the results in \cite{Dibitetto:2019odu,Balaguer:2023jei}, where it was shown that type IIB orientifold flux compactifications down to 6D, coupled to open strings may be mapped into non-chiral 6D gauged supergravities.

Our present setting is massive type IIA reductions on twisted tori with O$6$-planes and parallel D$6$-branes (see \cite{Andriot:2022bnb} for a recent analysis of this compactification class). The effective 4D description turns out to be a half-maximal gauged supergravity. These lower-dimensional theories with sixteen supercharges have been extensively studied in a bottom-up fashion by using the embedding tensor formalism \cite{deWit:2002vt} as an organizing principle \cite{Schon:2006kz}.

The 4D/10D dictionary relates fluxes to embedding tensor components (see \cite{Samtleben:2008pe} for a nice review). In this massive IIA setup the emebedding tensor/fluxes dictionary was analyzed in various works (see \emph{e.g.} \cite{Angelantonj:2003rq,Angelantonj:2003up,Angelantonj:2003zx,Roest:2009dq,DallAgata:2009wsi}). Later, in \cite{Dibitetto:2011gm}, this dictionary within the closed string sector was used in order to classify the SO$(3)$ invariant vacua of massive IIA on twisted tori, which are consistent with only one type of O$6$-planes and parallel D$6$-branes. The landscape of such vacua was organized in four discrete families of AdS extrema, one of which preserving a residual $\mathcal{N}=1$ supersymmetry.

The goal of this paper is to generalize the analysis carried out in \cite{Dibitetto:2011gm} by including open string dof's, such as dynamical brane position scalars and Wilson lines wrapping internal space, and their novel effects, such as Yang-Mills (YM)-like brane gauge groups and non-trivial flux associated with the internal YM field strength. We will show how the reduction of the non-Abelian WZ action for the D$6$-branes yields a modification of the bulk RR field strengths sourced by the open string dof's, just like in the case of the heterotic string, where this follows from the Green-Schwarz (GS) mechanism \cite{Green:1984sg} for anomaly cancellation. Indeed, the modifications we find may be heuristically obtained from the GS terms via a duality chain.

Subsequently, we present a detailed derivation of the scalar potential arising from dimensional reduction of the 10D bulk action plus the non-Abelian DBI brane action. This is then matched with the scalar potential of a suitable half-maximal gauged supergravity theory, including extra vector multiplets that take the open string dof's into account. It is worth mentioning that the full matching works out, if one expands the DBI at NLO in $\alpha'$.

Finally, the aforementioned scalar potential is studied, with the aim of finding novel extrema corresponding to string vacua featuring an interplay between open \& closed strings. We find new examples of AdS vacua, that are organized in a very similar way to their pure closed-string counterparts found in \cite{Dibitetto:2011gm}. By analyzing the supersymmetry conditions, we realize that only one family of vacua preserves minimal 4D supersymmetry, while all the other solutions are non-supersymmetric. Some of these, nevertheless, appear to be perturbatively stable.

It is worth noticing, though, that our vacua are crucially obtained by treating closed-string and open-string fluxes as competing effects. In some sense one might be led to think that this may require working at \emph{finite} $\alpha'$. Hence, it still remains to be assessed whether our construction is also physically reliable, besides its mathematical consistency. We conclude by studying the reliability of the novel solutions we found, by estimating the size of the higher-derivative corrections that would be present in this case and that we had been neglecting. A reliable perturbative corner for our solutions, where one has parametric control on such corrections, turns out to exist.

The paper is organized as follows. In Sec.~\ref{section:OrientifoldReduction} we review some relevant features of massive IIA twisted reductions with spacetime filling O$6$/D$6$ sources. In Sec. \ref{section:4DSugra} we illustrate the embedding tensor formulation of 4D $\mathcal{N}=4$ gauged supergravities, coupled to an arbitrary number of vector multiplets. In Sec.~\ref{section:Reduction} we go through the dimensional reduction of the 10D bulk action, as well as the non-Abelian WZ \& DBI D$6$-brane actions, and discuss the modified RR field strengths.
The vacua analysis is then carried out in Sec.~\ref{section:Vacua}, where we present our solutions to the field equations and discuss their mass spectra and supersymmetry.
In Sec.~\ref{section:PerturbativeControl} we address the issue of flux quantization, their perturbative control, and higher-derivative corrections.
Finally, some technical support material may be found in our appendices.  These contain detailed information concerning the democratic formulation of massive IIA supergravity, non-Abelian brane actions, and reductions thereof.

\section{Orientifold reduction of massive IIA supergravity with open strings}
\label{section:OrientifoldReduction}

In this section we describe the compactification of type IIA superstring effective action on twisted tori to 4D in the presence of D6-branes parallel to O6-planes, according to the following set-up:
\begin{align}
\mathrm{O}6 \ : \quad
\underbrace{\times \ | \ \times \ \times \ \times}_{\textrm{4D spacetime}}\ \  \underbrace{\underbrace{\times \ \times \ \times}_{y^a} \underbrace{\ - \ - \ -}_{y^i}}_{y^m} \ .    
\end{align}
Let us note that the tension of the O6-plane, which coincides with its charge, is given by \cite{Bergman:2001rp}
\begin{align}
T_{\text{O}6^{\pm}}=\mu_{\text{O}6^{\pm}}=\pm 2 \,T_{\text{D}6} \ ,
\end{align}
where $T_{\text{D}6}$ is the tension of the D$6$-brane and amounts to
\begin{align}
    T_{\text{D}6}=2\pi(2\pi\ell_s)^{-7} 
    \ ,
\end{align}
where $\ell_s$ is the string length. Henceforth we will consider the normalization $2\pi\ell_s =1$, except in Section \ref{section:PerturbativeControl}, where we will manifestly restore the dependence on $\ell_s$.

According to the above orientifold configuration in 10D, the orientifold involution on the internal coordinates is
\begin{align}    
\sigma_{\mathrm{O}6} \ : \left\{ y^a \to y^a \ , \ a=1,2,3 \ , \atop y^i \to -y^i  \ , \ i=4,5,6 \ ,\right.
\end{align}
where the internal longitudinal coordinates are represented by $a, b, c$ and the internal transverse coordinates are denoted by $i, j, k$. According to this configuration, let us assume the following compactification ansatz for the 10D metric:
\begin{equation} \label{eq:metric_Ansatz}
    ds^2_{(10)}=\, \tau^{-2} \, g_{\mu \nu} \, dx^{\mu} \, dx^{\nu} + \rho \,( \, \sigma^2 \, M_{ab} \, e^a \, e^b + \sigma^{-2} \, M_{ij} \, e^i \, e^j \,)  
\ ,
\end{equation}
where $g_{\mu\nu}$ is the 4D metric, $M_{ab}$ and $M_{ij}$ are two symmetric matrices which are elements of the $\left.\text{SL}(3)/\text{SO}(3)\right|_\text{L}$ and the $\left.\text{SL}(3)/\text{SO}(3)\right|_\text{R}$ scalar cosets spanned by the internal longitudinal and internal transverse components of the 10D metric, respectively. On the other hand, $e^m$ are the Maurer-Cartan 1-forms \cite{Scherk:1979zr}, which fulfill the equation
\begin{align}
    \dd e^m +\frac12 \,\omega_{np}{}^m \, e^n\wedge e^p =0 \ ,
\end{align}
where $\omega_{np}{}^m$ are the structure constants. They satisfy the Jacobi equation as an integrability condition,
\begin{align}
    \omega_{[mn}{}^r\, \omega_{p]r}{}^q=0
    \ ,
\end{align}
together with the unimodularity condition $\omega_{mn}{}^n=0$, as we are doing a compactification at the level of the action.

The lower dimensional effective theory consists of an ${\cal N}=4=D$ supergravity theory with global symmetry $\mathrm{SL}(2,\R)\times\mathrm{SO}(6,6+\mathfrak{N})$, and whose scalar coset is spanned by 
\begin{align}
\frac{\mathrm{SL}(2,\R)}{\mathrm{SO}(2)}\times\frac{ \mathrm{SO}(6,6+\mathfrak{N})}{\mathrm{SO}(6)\times \mathrm{SO}(6+\mathfrak{N})}    
\ .
\end{align}
This $(38+6\mathfrak{N})$-dimensional coset is filled out with the DOFs arising from the internal components of the supergravity and the brane fields. These scalars and the set of fluxes, which must be consistent with the O6 involution together with the combination of the fermionic number symmetry $(-1)^{F_L}$ and the worldsheet parity symmetry $\Omega$ are shown in Table \ref{tab:scalars}.

\begin{table}[t]
\centering
\begin{tabular}{|c|c|c|c|}
\hline
Type IIA Field   			  & $\sigma_{\mathrm{O}6}$  & $(-1)^{F_L}\Omega$ 	&	physical dof's\\ \hline \hline
$\Phi$         			      & $+$           & $+$     &		1          \\ \hline 
$B_{ai}$    			      & $-$           & $-$     &		9          \\ \hline
$C_i$        			      & $-$           & $-$     &		3          \\ \hline
$C_{abc}$   		          & $+$           & $+$     &		1          \\ \hline
$C_{aij}$     			      & $+$           & $+$     &		9          \\ \hline
$C_{abijk}$    			      & $-$           & $-$     &       3		   \\ \hline
$g_{ab}$                      & $+$           & $+$     &       6   	   \\ \hline
$g_{ij}$      			      & $+$           & $+$     &		6          \\ \hline
$Y^{Ii}$    				  & $-$           & $-$     &		$3\, \mathfrak{N}$          \\ \hline
$\mathcal{A}^I{}_{a}$    	  & $+$           & $+$     &		$3\, \mathfrak{N}$          \\ \hline
\end{tabular}
\begin{tabular}{|c|c|c|}
\hline
Type IIA Flux  			      & $\sigma_{\mathrm{O}6}$ & $(-)^{F_L}\Omega$ \\ \hline \hline
$\omega_{ab}{}^c$ 			  & $+$           & $+$               \\ \hline
$\omega_{ij}{}^c$ 			  & $+$           & $+$               \\ \hline
$\omega_{ai}{}^j$	  	 	  & $+$           & $+$               \\ \hline
$H_{ijk}$      			      & $-$           & $-$               \\ \hline
$H_{abi}$     			      & $-$           & $-$               \\ \hline
$F_{(0)}$    			      & $+$           & $+$               \\ \hline
$F_{ai}$      			      & $-$           & $-$               \\ \hline
$F_{abij}$     			      & $+$           & $+$               \\ \hline
$F_{abcijk}$     			  & $-$           & $-$               \\ \hline
$\mathcal{F}^I{}_{ab}$    	  & $+$           & $+$               \\ \hline
\end{tabular}
\caption{\emph{Propagating scalar DOF's (left) and fluxes (right) that are projected in by the $\mathbb{Z}_2$ truncation due to the presence of O6/D6 objects in type IIA compactifications down to 4D.}}
\label{tab:scalars}
\end{table}


Let us note that the structure constants of the Yang-Mills gauge group $G_{\text{YM}}$  associated to the stack of D6 branes, despite not being genuine fluxes, they consist of deformation parameters of the lower dimensional effective theory. This group, which will be of the form
\begin{align} \label{eq:GYM}
    G_{\text{YM}}
    =
    \left(
    \prod_a \text{U}(N_a)
    \right)
    \times
    \left(
    \prod_b \text{SO}(2N_b)
    \right)
    \times
    \left(
    \prod_c \text{USp}(2N_c)
    \right)
\ ,
\end{align}
is crucially constrained by the number of massless vector fields of the lower dimensional effective theory, $\mathfrak{N}$, which adds up to
\begin{align}
\mathfrak{N}
=
\sum_a N_a^2
+\sum_b N_b(2N_b-1)
+\sum_c N_c(2N_c+1)
\ .
\end{align}
Thus, the resulting non-semisimple gauge group of the 4D effective supergravity will consist of $G_{\text{YM}}\times G_{\text{ISO}}$, where $G_{\text{ISO}}$ corresponds to the gaugings associated to the isometries of the compactified bulk theory.

\subsection{SO(3) truncation and \texorpdfstring{${\cal N}=1$}{N=1} superpotential}
Many interesting type IIA vacua have beeen found in ${\cal N}=1$ compactifications. Interestingly, a part of them correspond to some specific truncations of the ${\cal N}=4$ effective supergravity theory obtained from the D6/O6 compactification. From a 10D viewpoint, this is carried out by compactifying the theory in the presence of extended objects, together with further consistent truncations arising from orbifold projections and/or preserving a specific subgroup of the global symmetry group.

In this work, we will consider an SO(3) truncation, namely we will just keep the fluxes and scalar fields that transform as singlet under the SO(3) subgroup of $\mathrm{SO}(6,6+\mathfrak{N})$. Upon this truncation, the theory reduces to
\begin{align}
\frac{\mathrm{SL}(2, \R)}{\mathrm{SO}(2)}\times\frac{ \mathrm{SO}\left(2,2+\frac{\mathfrak{N}}{3}\right)}{\mathrm{SO}(2)\times \mathrm{SO}\left(2+\frac{\mathfrak{N}}{3}\right)}    
\ ,
\end{align}
whose dimension is $6+2\mathfrak{N}/3$. For the simplest case $\mathfrak{N}=3$, the 8-dimensional scalar coset is spanned by the following closed string complex moduli
\begin{align}
S=&\ \chi+ie^{-\phi}
\ ,
&
T=&\ \chi_1+ie^{-\varphi_1}
\ ,
&
U=&\ \chi_2+ie^{-\varphi_2}
\ ,
\end{align}
together with two open string scalar fields, ${\cal A}$ and $Y$, such that $\{ {\cal A}^I{}_a,\, {Y}^{Ii}\}=\{\delta^I_a \, {\cal A},\, \delta^{Ii} Y\}$. The axions of the complex scalars parametrize the following scalar fields:
\footnote{
Let us stress that the SO(3) truncation defines the diagonal group of $\text{SO}(3)_a\times \text{SO}(3)_i \times \text{SO}(3)_I$ (see \cite{Dibitetto:2011gm} for further details).
}
\begin{align} \label{eq:dictionary_chi}
C_{abc}=&\ \epsilon_{abc} \chi
\ ,
&
C_{aij}=&\ \epsilon_{aij} \chi_1
\ ,
&
B_{ai}=&\ \delta_{ai} \chi_2
\ ,
\end{align}
whereas the dilatons are written in terms of the universal moduli as
\begin{align} \label{eq:dictionary_phi}
    e^{\phi}
    =&\ 
    \frac{1}{\tau \sigma^3}
    \ , 
    &
    e^{\varphi_1}
    =&\ 
    \frac{\sigma}{\tau}
    \ , 
    &
    e^{\varphi_2}
    =&\ 
    \frac{1}{\rho} 
    \ .
\end{align}

To understand the existence of an ${\cal N}=1$ formulation, let us consider the R-symmetry group of $\mathcal{N}=4$ theory $\text{SU}(4)\sim\text{SO}(6)$. Interestingly, when studying the splitting of the fundamental representation under the action of $\text{SO}(3)$, $\mathbf{4}\to \mathbf{1}\oplus  \mathbf{3}$, we note that an $\mathcal{N}=1$ structure still is possible due to the presence of the singlet. In particular, the scalar potential of the closed string sector 
\footnote{It would be interesting to have a complete ${\cal N}=1$ open-string superpotential \cite{Escobar:2018tiu} that accounts for our set of fluxes.} is given, up to $\N=4$ quadratic constraints, by
\begin{align}
    V
    =
    e^{K}\left(\sum_\Phi K^{\Phi\bar{\Phi}}|D_\Phi W|^2-3|W|^2\right)
    \ ,
\end{align}
where $\Phi\equiv (S,T,U)$, $K\equiv K(\Phi,\bar\Phi)$, is the K\"ahler potential,
\begin{align}
    K(\Phi, \bar{\Phi})=-\log(-i(S-\bar{S}))-3\log(-i(T-\bar{T}))-3\log(-i(U-\bar{U}))
    \ , 
\end{align}
and $W=W(\Phi)$ is the holomorphic superpotential
\begin{align}
    W(\Phi)
    =
    P_F-P_H\, S+3P_Q\, T
    \ ,
\end{align}
with the polynomials $P_F, P_H$ and $P_Q$ defined as follows:
\begin{align}
    & P_F
    =
    a_0-3\, a_1\, U+3\, a_2\, U^2-a_3\, U^3 \ ,\\
    & P_H
    =
    b_0-3\, b_1\, U\ ,\\
    & P_Q
    =
    c_0+(2c_1-\bar{c}_1)\, U \ .
\end{align}
Here, $K^{\Phi\bar{\Phi}}$  represents the inverse of the  K\"ahler metric $K_{I\bar{J}}=\frac{\partial K}{\partial \Phi^I\partial {\bar{\Phi}}^{\bar{J}}}$ and the K\"ahler derivative is $D_\Phi W=\frac{\partial W}{\partial \Phi}+\frac{\partial K}{\partial \Phi}W$.

The correspondence between the type IIA fluxes and the fluxes in the superpotential formulation is summarized in Table \ref{tab:fluxes-parametrization}.

\begin{table}[t]
\centering
\begin{tabular}{|c|c|c|}
\hline
	Type IIA  	  		  				& fluxes  & Parametrization \\ \hline \hline
	$F_{aibjck}$ 		    			& $a_0$   & $F_{aibjck}= \varepsilon_{abc} \, \varepsilon_{ijk} \, a_0$ 
 \\ \hline
	$F_{aibj}$     						& $a_1$   &  $F_{aibj}= \varepsilon_{abc}\, \varepsilon_{ijk} \,\delta^{ck} \, a_1$
 \\ \hline
    $F_{ai}$     	  					& $a_2$   &   $F_{ai}= \delta_{ai} \, a_2 $
    \\ \hline
	$F_{0}$     	  					& $a_3$   &  $F_{0}=a_3$
 \\ \hline \hline

	$H_{ijk}$      	  					& $b_0$   &  $H_{ijk}=\varepsilon_{ijk} \, b_0$ \\ \hline 
	$\omega_{ij}{}^c$ 	    			& $b_1$   &   $\omega_{ij}{}^c=\varepsilon_{ijd} \, \delta^{cd} \, b_1$\\ \hline \hline
	
	$H_{abk}$ 	    					& $c_0$   &   $H_{abk}=\varepsilon_{abk} \,  c_0$\\ \hline 
	$\omega_{ka}{}^j=\omega_{bk}{}^i$   & $c_1$   &  $\omega_{ka}{}^j=\varepsilon_{kal} \, \delta^{lj} \, c_1 $ \\ \hline 
	$\omega_{bc}{}^a$ 	 	    		& $\bar{c}_1$  & $\omega_{bc}{}^a=\varepsilon_{bcd} \, \delta^{ad} \, \bar{c}_1 $   \\ \hline \hline

    $\mathcal{F}^K{}_{ab}$   	    	& $g_{0}$ &  $\tensor{\mathcal{F}}{^K_{ab}}=\tensor{\varepsilon}{_{abc}} \; \tensor{\delta}{^{cK}} g_0$  \\ \hline 	
	$g_{IJ}{}^K$ 	    					& $g_{1}$ &  $g_{IJ}{}^K=\epsilon_{IJL}{} \delta^{LK} g_{1}$   \\ \hline 
\end{tabular}
\caption{\emph{Parametrization of $\text{SO}(3)$ invariant type IIA fluxes.}}
\label{tab:fluxes-parametrization}
\end{table}

\section{\texorpdfstring{$\mathcal{N}=4$}{N=4} supergravity with vector multiplets} \label{section:4DSugra}
Reduction of massive type IIA supergravity on 6-dimensional twisted tori with orientifold admits a 4D gauged supergravity description \cite{DallAgata:2009wsi}: the supergravity theory we are interested in is an half-maximal supergravity, whose detailed description can be found in \cite{Schon:2006kz}. 

In principle, we can consider an  $\mathcal{N}=4$ supergravity theory where the gravity multiplet is coupled to an arbitrary number $n$ of vector multiplets. If we are interested in matching the lower dimensional theory with the closed string sector only, then we need to have 6 vector multiplets, as in \cite{Dibitetto:2011gm}.
The possibility of coupling to $\mathfrak{N}$ extra vector multiplets besides the $6$ associated to the closed string modes, as it is described in \cite{Danielsson:2017max},  allows to capture also the dynamics of open strings living on D6 branes parallel to the O6 plane. In this case the $\mathcal{N}=4$ theory enjoys a global symmetry
\begin{equation}
    G_{\textup{global}}= \mathrm{SL}(2,\mathbb{R}) \times \mathrm{SO}(6,6+\mathfrak{N}) \;.
\end{equation}
The bosonic degrees of freedom of the theory amount to the graviton, $12+ \mathfrak{N}$ vectors and $38+6 \mathfrak{N }$ scalars, which can be mapped to those coming from type IIA supergravity after the orientifold projection and dimensional reduction are performed, as detailed in Table \ref{tab:scalars}.

The scalar degrees of freedom of the theory, in particular, can be described through the coset manifold
\begin{equation}
    \mathcal{M}_{\textup{scalar}} =\underbrace{\frac{\mathrm{SL}(2,\mathbb{R})}{\mathrm{SO}(2)}}_{M_{\alpha \beta}} \times \underbrace{\frac{\mathrm{SO}(6,6+ \mathfrak{N})}{\mathrm{SO}(6)\times \mathrm{SO}(6+ \mathfrak{N})}}_{M_{MN}} \; ,
\end{equation}
where the $6 \, \mathfrak{N}$ extra scalar modes are to be interpreted as corresponding to $3 \, \mathfrak{N}$ position moduli of the D6 branes in the transverse directions $Y^{Ii}$ plus $3 \, \mathfrak{N}$ degrees of freedom corresponding to the internal components of the worldvolume gauge fields $\mathcal{A}^I_a$ (with the notation $I=1,\ldots \mathfrak{N}$). Coset representatives are denoted with $M_{\alpha \beta}$, where $\alpha,\, \beta= (+,-)$ are $\mathrm{SL}(2,\mathbb{R})$ indices, which can be raised or lowered by $\varepsilon_{\alpha \beta}=\varepsilon^{\alpha \beta}$ (our convention is that $\varepsilon_{+-}=1$), and $M_{MN}$, where $M,\,N$ are $\mathrm{SO}(6, 6+\mathfrak{N})$ indices. Here we adopt a light-cone basis along the $\mathrm{SO}(6, 6)$ coordinates, but a Cartesian basis in the extra $\mathrm{SO}(\mathfrak{N})$ directions, so that the metric raising and lowering the $\mathrm{SO}(6, 6+\mathfrak{N})$ indices is
\begin{equation} \label{eq:Metric}
    \eta_{MN}= \eta^{MN}= \begin{pmatrix}
        \mathbb{O}_6 & \mathbb{I}_6 & \mathbb{O}_{6,\mathfrak{N}} \\
        \mathbb{I}_6 & \mathbb{O}_6 & \mathbb{O}_{6,\mathfrak{N}} \\
        \mathbb{O}_{\mathfrak{N},6} & \mathbb{O}_{\mathfrak{N},6} & \mathbb{I}_{\mathfrak{N}} \\
    \end{pmatrix}\;.
\end{equation}
In the following, we will always consider that the global $\mathrm{SO}(6, 6+\mathfrak{N})$ indices are split as $M \equiv (m, \bar{m}, I )$, with $m, \bar{m}= 1,\ldots 6$, $I=1,\ldots \mathfrak{N}$. Furthermore, the indices  $m, \bar{m}$ can be split as $m = (a,i)$, $\bar{m}=(\bar{a},\bar{i})$, where $a,i,\bar{a} , \bar{i} = 1,2,3$, in analogy with the splitting of the internal coordinates.  
The first factor in our coset can be parametrized as
\begin{equation} \label{eq:Mab_Parametrization}
    M_{\alpha \beta}= e^{\phi}\, \begin{pmatrix}
     \chi ^2 + e^{-2\phi} & \chi \\
     \chi & 0 \\
    \end{pmatrix} \;.
\end{equation}
An explicit parametrization of the matrix $M_{MN}$, instead, can be obtained starting from two $6 \times 6$ matrices, which we call $G$ and $B$, respectively symmetric and antisymmetric. If we also define an $\mathfrak{N} \times 6 $ matrix containing the scalar degrees of freedom from the open string sector
\begin{equation}
    \tensor{A}{^I_m} \equiv \big( \tensor{\mathcal{A}}{^I_a}, Y^{Ii} \big) \;,
\end{equation}
then we can introduce $C=B+ \frac{1}{2} A^T A$ and write the full $M_{MN}$ as 
\begin{equation} \label{eq:M_MN_Parametrization}
    M_{MN}= \begin{pmatrix}
        G^{-1} & -G^{-1} C & -G^{-1} A^{T} \\
        -C^{T} G^{-1} & \quad G+C^{T} G^{-1} C + A^{T} A \quad & C^{T} G^{-1} A^{T} + A^{T} \\
        -A G^{-1} & A G^{-1} C + A & \mathbb{I} + A G^{-1} A^{T} \\
    \end{pmatrix} \;.
\end{equation}
The matrix $M_{MN}$ can be obtained, equivalently, from a vielbein $\tensor{\mathcal{V}}{_M^{\underline{M}}}$ (where $\underline{M}$ is a local $\mathrm{SO}(6) \times \mathrm{SO}(6+\mathfrak{N})$ index with a splitting analogous to $M$), such that $M = \mathcal{V}^T \, \mathcal{V}$.  In order to be consistent with our parametrization \eqref{eq:M_MN_Parametrization}, we have to choose this vielbein to be
\begin{equation}
    \tensor{\mathcal{V}}{_M^{\underline{N}}}= \begin{pmatrix}
        E^{T} & 0  & 0\\
        -C^{T} E^{T} & E^{-1} & A^{T} \\
        -A E^{T } & 0 & \mathbb{I}
    \end{pmatrix}\ ,
\end{equation}
with $E^{T} E = G^{-1}$. 

The embedding of the gauge group of the theory into the global symmetry group can be parametrized by two tensors $\xi_{\alpha M}$ and $f_{\alpha MNP}=f_{\alpha [MNP]}$ \cite{Schon:2006kz}, but for the cases we will consider we will always have $\xi_{\alpha M}=0$, then in the rest of this paper we will always consider the embedding tensor to be given only by $f_{\alpha MNP}$. The quadratic constraints it has to satisfy in order to ensure closure of the gauge algebra are
\begin{equation}
    f_{\alpha R [MN} \tensor{f}{_{\beta PQ]}^R}=0 \;, \qquad \varepsilon^{\alpha \beta}  f_{\alpha MNR} \tensor{f}{_{\beta PQ}^R}=0 \;.
\label{eq:QC}
\end{equation}
The scalar potential of the theory, which is induced by the gauging, can be expressed in terms of the embedding tensor and the scalar fields as
\begin{equation} \label{eq:GaugedSugraPotential}
\begin{split}
    V=& \;\frac{1}{64} f_{\alpha MNP} f_{\beta QRS} M^{\alpha \beta} \Big[ \frac{1}{3} M^{MQ} M^{NR} M^{PS} + \Big( \frac{2}{3} \eta^{MQ} - M^{MQ}\Big) \eta^{NR} \eta^{PS} \Big]  \\
    &- \frac{1}{144} f_{\alpha MNP } f_{\alpha QRS} \varepsilon^{\alpha \beta} M^{MNPQRS} \;.
\end{split} 
\end{equation}
Here the antisymmetric tensor $M^{MNPQRS}$ is defined (after raising the indices with the metric \eqref{eq:Metric}) as
\begin{equation}
    M_{MNPQRS} \equiv \varepsilon_{\underline{m} \underline{n} \underline{p} \underline{q} \underline{r} \underline{s}} \tensor{\mathring{\mathcal{V}}}{_M^{\underline{m}}} \tensor{\mathring{\mathcal{V}}}{_N^{\underline{n}}} \tensor{\mathring{\mathcal{V}}}{_P^{\underline{p}}} \tensor{\mathring{\mathcal{V}}}{_Q^{\underline{q}}} \tensor{\mathring{\mathcal{V}}}{_R^{\underline{r}}} \tensor{\mathring{\mathcal{V}}}{_S^{\underline{s}}} \;,
\end{equation}
where $\tensor{\mathring{\mathcal{V}}}{_M^{\underline{M}}}=\tensor{\mathcal{V}}{_M^{\underline{N}}} \tensor{R}{_{\underline{N}}^{\underline{M}}}$ is a vielbein where the local index has been rotated to Cartesian coordinates through a matrix
\begin{equation}
    R \equiv \frac{1}{\sqrt{2}} \begin{pmatrix}
        -\mathbb{I}_6 & \mathbb{I}_6 & \mathbb{O}_{6,\mathfrak{N}} \\
        \mathbb{I}_6 & \mathbb{I}_6 & \mathbb{O}_{6,\mathfrak{N}} \\
        \mathbb{O}_{\mathfrak{N},6} & \mathbb{O}_{\mathfrak{N},6} & \sqrt{2} \;\mathbb{I}_{\mathfrak{N}} \\
    \end{pmatrix} \;.
\end{equation}

\subsection{The $\mathrm{SO}(3)$ truncation}

In analogy with \cite{Dibitetto:2011gm}, we consider an $\mathrm{SO}(3)$ truncation of the full supergravity theory we are studying. Namely, we define an $\mathrm{SO}(3)$ subgroup of the global symmetry group $\mathrm{SO}(6, 6+\mathfrak{N})$ and we retain only the scalar fields and embedding tensor components that transform as singlets with respect to it. Setting to zero all non-singlet scalars is allowed since their equation of motion are consistently satisfied (they can never be sourced by $\mathrm{SO}(3)$ singlets). 

The truncation makes the scalar sector of the theory reduce to
\begin{equation}
    \mathcal{M}_{\mathrm{SO}(3)} =\frac{\mathrm{SL}(2,\mathbb{R})}{\mathrm{SO}(2)} \times \frac{\mathrm{SO} \Big(2,2+ \frac{\mathfrak{N}}{3}\Big)}{\mathrm{SO}(2)\times \mathrm{SO}\Big(2+ \frac{\mathfrak{N}}{3}\Big)}\; ,
\end{equation}
which in the parametrization \eqref{eq:M_MN_Parametrization} of the $M_{MN}$ matrix corresponds to defining the $G$ and $B$ matrices in the following way:
\begin{equation}
     G= e^{\varphi_2 - \varphi_1}\, \begin{pmatrix}
     \chi_2 ^2 + e^{-2\varphi_2} & -\chi_2 \\
     -\chi_2 & 1 \\
    \end{pmatrix} \otimes \mathbb{I}_3\;, \qquad B=  \begin{pmatrix}
     0 & \chi_1 \\
     -\chi_1 & 0 \\
    \end{pmatrix} \otimes \mathbb{I}_3\;.
\end{equation}
Hence, also considering the parametrization \eqref{eq:Mab_Parametrization}, the scalar degrees of freedom surviving in the truncated theory are three dilatons $(\phi, \varphi_1, \varphi_2 )$ and three axions $(\chi, \chi_1, \chi_2 )$, plus the scalars coming from the extra $\mathfrak{N}$ vector multiplets.

The correspondence between the three dilatons $\phi$, $\varphi_1$, $\varphi_2$ surviving the $\mathrm{SO}(3)$ truncation and the universal moduli $\rho$, $\sigma$, $\tau$ appearing in the parametrization of the ten-dimensional metric \eqref{eq:metric_Ansatz}, which after the $\mathrm{SO}(3)$ truncation takes the form
\begin{equation}
     \dd s^2_{(10)}=\, \tau^{-2} \, g_{\mu \nu} \, \dd x^{\mu} \, \dd x^{\nu} + \rho \,( \, \sigma^2 \, \delta_{ab} \, e^a \, e^b + \sigma^{-2} \, \delta_{ij} \, e^i \, e^j \,)   
\end{equation}
is given by
\begin{equation}
    e^{\phi}=\frac{1}{\tau \sigma^3}, \qquad e^{\varphi_1}=\frac{\sigma}{\tau}, \qquad e^{\varphi_2}=\frac{1}{\rho},
\end{equation}
or, inverting the relations,
\begin{equation}
    \tau=e^{\frac{1}{4}(-\phi-3\varphi_1)}. \qquad \rho=e^{-\varphi_2}, \qquad \sigma=e^{\frac{1}{4}(-\phi+\varphi_1)}.
\end{equation}
This matching of the scalar degrees of freedom allows us to determine, for any term in the action of the type IIA supergravity having its own scaling with respect to the universal moduli, the corresponding scaling with respect to the gauged supergravity scalars. The results, reported in table \ref{Table:Scaling_of_fluxes}, will be useful in order to complete the dictionary between fluxes and embedding tensor components and also to understand the structure of modified field strengths in section \ref{section:Reduction}. 

\begin{table}[H]
\centering
\renewcommand{\arraystretch}{1.3}
\begin{tabular}{|c|c|c|}
    \hline
         Fluxes & Scaling w.r.t. $\phi$, $\varphi_1$, $\varphi_2$ & Scaling w.r.t. $\rho$, $\sigma$, $\tau$ \\
    \hline \hline
    
        $\abs{F_6}^2$ & $e^{\phi + 3 \varphi_1 +3 \varphi_2}$ & $\tau^{-4} \rho^{-3}$ \\[3pt] \hline
        $\abs{F_4}^2$ & $e^{\phi + 3 \varphi_1 + \varphi_2}$ & $\tau^{-4} \rho^{-1}$ \\[3pt] \hline 
        $\abs{F_2}^2$ & $e^{\phi + 3 \varphi_1 - \varphi_2}$ & $\tau^{-4} \rho$ \\[3pt] \hline
        $\abs{F_0}^2$ & $e^{\phi + 3 \varphi_1 -3 \varphi_2}$ & $\tau^{-4} \rho^{3}$ \\[3pt] \hline
        $\abs{H_{ijk}}^2$ & $e^{-\phi + 3 \varphi_1 +3 \varphi_2}$ & $\tau^{-2} \rho^{-3} \sigma^{6}$ \\ [3pt] \hline
        $\abs{\tensor{\omega}{_{ij}^c}}^2$ & $e^{-\phi + 3 \varphi_1 + \varphi_2}$ & $\tau^{-2} \rho^{-1} \sigma^{6}$ \\[3pt] \hline
        $\abs{H_{abk}}^2$ & $e^{\phi +  \varphi_1 +3 \varphi_2}$ & $\tau^{-2} \rho^{-3} \sigma^{-2}$ \\[3pt] \hline
        $\abs{\tensor{\omega}{_{ka}^j}}^2$, $\abs{\tensor{\omega}{_{bc}^a}}^2$ & $e^{\phi + \varphi_1 + \varphi_2}$ & $\tau^{-2} \rho^{-1} \sigma^{-2}$ \\[3pt] \hline
        $\tensor{g}{_{KL}^I} \tensor{g}{_{MN}^J}$ & $e^{\phi + 2(\varphi_1 - \varphi_2)}$ & $\tau^{-3} \rho^{2} \sigma^{-1}$ \\[3pt] \hline
        $\abs{\tensor{\mathcal{F}}{^I_{ab}}}^2$ & $e^{\phi + 2(\varphi_1 + \varphi_2)}$ & $\tau^{-3} \rho^{-2} \sigma^{-1}$ \\[3pt] \hline
   
\end{tabular}
\caption{\emph{Scaling of the fluxes with respect to the dilatons $\phi$, $\varphi_1$, $\varphi_2$ and the universal moduli $\rho$, $\sigma$, $\tau$.}}
\label{Table:Scaling_of_fluxes}
\end{table}

As far as the embedding tensor is concerned, in order to be consistent with the $\mathrm{SO}(3)$ truncation it can be factorized as
\begin{equation} \label{eq:def_Lambda}
    \tensor{f}{_{\alpha MNP}}= \tensor{\Lambda}{_{\alpha ABC}} \;\tensor{\varepsilon}{_{xyz}} \;,
\end{equation}
(in analogy with \cite{Dibitetto:2011gm}) where the tensor $\tensor{\Lambda}{_{\alpha ABC}}$ is completely symmetric in the indices $A,\, B, \, C$ running from $1$ to $\displaystyle{\frac{12+ \mathfrak{N}}{3}}$. The quadratic constraints for $f_{\alpha MNP}$, in this case, translate into the following equations for $\Lambda_{\alpha ABC}$
\begin{equation} \label{eq:quadraticconstraints}
    \tensor{\Lambda}{_{(\alpha A [ B}^C} \tensor{\Lambda}{_{\beta) D ] EC}}=0  \;, \qquad \varepsilon^{\alpha \beta} \tensor{\Lambda}{_{\alpha AB}^C} \tensor{\Lambda}{_{\beta DEC}}=0 \;.
\end{equation}

Since we want our gauged supergravity in 4 dimensions to match the dimensional reduction of type IIA supergravity, we cannot admit the most general form for the embedding tensor. Instead, we have to turn on only those components that can be associated to a flux in the higher dimensional theory; in particular, if we consider geometric fluxes coming from the closed string sector (but neglect the possibility of non-geometric ones), the corresponding components of the embedding tensor are listed in table \ref{Table:Dictionary_fluxes_embedding_closedsector} (the dictionary is reported both for general fluxes-embedding tensors and in the case of $\mathrm{SO}(3)$ truncation).

\begin{table}[H]
\centering
\renewcommand{\arraystretch}{1.15}
\(\begin{array}{|c|c|c|c|}
    \hline
        \quad \text{Type IIA} \quad & \quad \text{Fluxes} \quad & \quad \mathrm{SO}(6,6+\mathfrak{N}) \quad & \quad \mathrm{SO}(2,2+\mathfrak{N}/3) \quad \\[3pt]
    \hline \hline
        F_{aibjck} & a_0 & -f_{+ \bar{a} \bar{b} \bar{c}} & -\Lambda_{+333} \\[4pt] \hline
        F_{aibj} & a_1 & f_{+ \bar{a} \bar{b} \bar{k}}& \Lambda_{+334} \\[4pt] \hline
        F_{ai} & a_2 & -f_{+ \bar{a} \bar{j} \bar{k}} & -\Lambda_{+344} \\[4pt] \hline
        F_{(0)} & a_3 & f_{+ \bar{i} \bar{j} \bar{k}} & \Lambda_{+444} \\[4pt] \hline
        H_{ijk} &  b_0 & -f_{- \bar{a} \bar{b} \bar{c}}& -\Lambda_{-333}\\[4pt] \hline
        H_{abk} &  c_0 & f_{+ \bar{a} \bar{b} k}& \Lambda_{+233} \\[4pt] \hline
        \tensor{\omega}{_{ij}^c} & b_1 & f_{- \bar{a} \bar{b} \bar{k}} & \Lambda_{-334} \\[4pt] \hline
        \tensor{\omega}{_{ka}^j}=\tensor{\omega}{_{bk}^i} & c_1 & f_{+ \bar{a} \bar{j} k}=f_{+ \bar{i} \bar{b} k} & \Lambda_{+234} \\[4pt] \hline
        \tensor{\omega}{_{bc}^a} & \bar{c}_1 & f_{+ a \bar{b} \bar{c}} & \Lambda_{+133} \\[4pt] \hline
    \end{array}\)   
\caption{\emph{Dictionary between embedding tensor components and geometric fluxes of type IIA supergravity in the $\mathrm{SO}(3)$ truncation.}}
\label{Table:Dictionary_fluxes_embedding_closedsector}
\end{table}

\subsection{Our setup: coupling to $\mathfrak{N}=3$ extra vector multiplets}
We focus on the particular case when our supergravity theory is coupled to $\mathfrak{N}=3$ extra vector multiplets, so that the scalar fields parametrize the coset
\begin{equation}
    \frac{\mathrm{SL}(2,\mathbb{R})}{\mathrm{SO}(2)} \times \frac{\mathrm{SO}(6,9)}{\mathrm{SO}(6)\times \mathrm{SO}(9)} \; ,
\end{equation}
where the $18$ extra scalar modes are $(\mathcal{A}^I_a \,, \, Y^{Ii})$, with $I=1,2,3$. If we perform the $\mathrm{SO}(3)$ truncation, i.e. keep only the fields that transform as singlets under the $\mathrm{SO}(3)$ subgroup of the global symmetry group $\mathrm{SL}(2,\mathbb{R}) \times \mathrm{SO}(6,9)$, we are only left with the $6$ scalar degrees of freedom coming from the closed string sector, while the open string ones can be parametrized as
\begin{equation} \label{eq:openaxions}
    \Big(\mathcal{A}^I_a \,, \, Y^{Ii}\Big)= \Big(\mathcal{A} \, \delta^I_a \,, \, Y\, \delta^{Ii} \Big) \;.
\end{equation}

If we want to parametrize the embedding tensor as in \eqref{eq:def_Lambda}, in the case of $\mathfrak{N}=3$ the indices $A,\, B, \, C$ of the $\Lambda$ tensor run from $1$ to $5$. The correspondence between open string fluxes and embedding tensor components is
\begin{equation} \label{eq:openstring}
    \begin{split}
        & \tensor{\mathcal{F}}{^K_{ab}}=\tensor{\varepsilon}{_{abc}} \; \tensor{\delta}{^{cK}} g_0 \quad , \qquad  \ \ \tensor{\Lambda}{_{+ 3 3 5}}=  \;g_0 \quad, \\
        & \tensor{g}{_{IJ}^K}\,= \tensor{\varepsilon}{_{IJL}} \; \tensor{\delta}{^{LK}}\,g_1 \quad , \qquad  \tensor{\Lambda}{_{+ 5 5 5}}= \;g_1 \quad. \\
    \end{split}
\end{equation}

Similar to the O7/D7 system in type IIB \cite{Camara:2005dc,Balaguer:2023jei}, the quadratic constraints that we have to impose on the embedding tensor to ensure consistency of the gauging give rise to the following relation between the open string fluxes
\begin{align}
g_{IJ}{}^K  \, {\cal F}^I{}_{ab}=0
\ , 
\label{eq:QC-nonsemisimple}
\end{align}
which in this case translates to
\begin{equation}
\label{QC_open_N3}
    g_0\,g_1 =0 \; .
\end{equation}
Furthermore, they imply three constraints involving the fluxes from closed string sector, \emph{i.e.} (in agreement with \cite{Dibitetto:2011gm})
\begin{equation}
    c_1 (c_1-\bar{c}_1) =0 \;, \quad b_1 (c_1-\bar{c}_1) =0 \;, \quad a_3 \, c_0 + 2 \, a_2 \, c_1 -a_2 \, \bar{c}_1 =0 \;, 
\end{equation}
which may be understood as requirements on the absence of SUSY breaking sources such as KK$5$ monopoles (the first two constraints), and D$6$ branes wrapping $ajk$ cycles (the last one).

\section{10-dimensional supergravity action and reduction thereof} \label{section:Reduction}
In this section we analyse the dimensional reduction of the type IIA supergravity action in 10 dimensions in presence of D$6$-branes. Our purpose is to extract the contributions of each term of the action to the final expression of the scalar potential in the lower-dimensional theory, in order to compare them with the full 4D potential we get from the gauged supergravity description.

\subsection{Bulk action}
The dimensional reduction of the Einstein-Hilbert action in $10$ dimensions
\begin{equation}
\label{Einstein_action_10D}
    \int \dd ^{10} x \; \sqrt{-g^{(10)}} \; e^{-2 \Phi} \; \mathcal{R}^{(10)}
\end{equation}
can be performed by adopting the following parametrization for the metric
\begin{equation} \label{eq:compactmetric}
\begin{split}
    ds^2_{(10)}=&\, \tau ^{-2} \, ds^2_{(4)} + \rho \, ds^2_{(6)}\\
    =&\, \tau^{-2} \, g_{\mu \nu} \, dx^{\mu} \otimes dx^{\nu} + \rho  \, M_{mn} \, e^m \otimes e^n \\
    =&\, \tau^{-2} \, g_{\mu \nu} \, dx^{\mu} \otimes dx^{\nu} + \rho \,( \, \sigma^2 \, M_{ab} \, e^a \otimes e^b + \sigma^{-2} \, M_{ij} \, e^i \otimes e^j \,)  \;, 
\end{split}    
\end{equation}
where $\rho$ is the universal modulus allowing to normalize the internal metric $g^{(6)}$ as
\begin{equation}
    \int \dd ^{6} y \; \sqrt{g^{(6)}} = 1 \;,
\end{equation}
while $\tau$ is related to $\rho$ and to the dilaton $\Phi$ by the following constraint
\begin{equation} \label{eq:dilatonscaling}
    e^{2\Phi} = \tau^{-2} \rho^{3} \;,
\end{equation}
in such a way that the gravity term of the action in $4$ dimensions can be written in the Einstein frame. The modulus $\sigma$, on the other hand, is related to the relative scaling between the internal directions that belong to the worldvolume of the O6 plane and the transverse ones. If we rewrite the Ricci scalar $\mathcal{R}^{(10)}$ and the determinant of the metric $g^{(10)}$ respectively as
\begin{equation}
    \tau^2 \mathcal{R}^{(4)} + \rho^{-1} \mathcal{R}^{(6)} \; , \qquad \tau^{-4} \rho^{3} \sqrt{-g^{(4)}} \sqrt{g^{(6)}}\; , 
\end{equation}
then the dimensional reduction of \eqref{Einstein_action_10D} is
\begin{equation}
    \begin{split}
        & \int \dd ^4 x \sqrt{-g^{(4)}} \; \bigg( \tau^{-2} \rho^{3} e^{-2\Phi} \mathcal{R}^{(4)} + \tau^{-4} \rho^{2} e^{-2\Phi} \mathcal{R}^{(6)} \bigg) \\
        = & \int \dd ^4 x \sqrt{-g^{(4)}} \; \bigg( \mathcal{R}^{(4)} + \underbrace{\tau^{-2} \rho^{-1} \mathcal{R}^{(6)}}_{\displaystyle{-V_{\omega}}} \bigg)  \quad .\\ 
    \end{split}
\end{equation}
Here the first term is precisely the gravity action in $4$ dimensions, while the second one, which we have named as $-V_{\omega}$, contributes to the scalar potential. The internal part of the Ricci scalar $\mathcal{R}^{(6)}$ can be expressed in terms of the internal metric $M_{mn}$ and the metric flux $\tensor{\omega}{_{mn}^p}$ as
\begin{equation}
    \mathcal{R}^{(6)}= -\frac{1}{4} M_{mq} M^{nr} M^{ps} \tensor{\omega}{_{np}^q} \tensor{\omega}{_{rs}^m}-\frac{1}{2} M^{np} \tensor{\omega}{_{mn}^q} \tensor{\omega}{_{qp}^m} \;.
\end{equation}
In particular, if we parametrize both the metric and the metric flux according to the $\mathrm{SO}(3)$ truncation, we obtain that in our setup the contribution to the scalar potential from the Einstein-Hilbert action is
\begin{equation} \label{eq:Vomega}
V_{\omega}=-\frac{3}{2}\,\tau^{-2} \rho^{-1}\big(\, \bar{c}_1^2 \,\sigma^{-2} + 4 \, b_1 \, c_1 \, \sigma^2 - b_1^2 \, \sigma^6 \, \big) \;.
\end{equation}
On the other hand, in the context of fluxes, the contribution to the scalar potential from the $H_{(3)}$ flux is
\begin{align}
    \int \dd^{10}x\, \sqrt{-g^{(10)}}\left(
	-\frac{1}{12}\, e^{-2\Phi}\, |H_{(3)}|^2
	\right)
\to
    \int \mathrm{d}^4 x\sqrt{-g^{(4)}}\left(-\frac{1}{12}H_{mnp}H^{mnp}\tau^{-2}\rho^{-3} \right),
\end{align}
where the contraction with the indices has been done with the internal metric $g^{(6)}$. Taking into account the $\mathbb{Z}_2$ truncation due to the presence of the O6-planes, and the SO(3) truncation, we can expand it as
\begin{align} \label{eq:VH3}
    V_{H_{3}}
    =\frac{1}{12}\tau^{-2}\rho^{-3}\left(b_0^2\, \sigma^6+3\,  c_0^2\, \sigma^{-2}\right) \ .
\end{align}
Regarding the R--R sector we have
\begin{align}
    \int \dd^{10} x \, \sqrt{-G}\left(
	-\frac{1}{2p!}\,  |F_{(p)}|^2
	\right)
\to
    \int \mathrm{d}^4 x\sqrt{-g^{(4)}}\left(-\frac{1}{2p!}F_{m_1\ldots m_p}F^{m_1\ldots m_p}\, \tau^{-4}\rho^{3-p} \right) \ .
\end{align}
So, the contribution to the scalar potential of each field strength is
\begin{eqnarray}
V_{F_p} &=& \frac{1}{2p!}F_{m_1\ldots m_p}F^{m_1\ldots m_p} \tau^{-4}\rho^{3-p}\ .
\end{eqnarray}
The expansion of this term can be explicitly done for each $p$-form. Due to the $\mathbb{Z}_2$ truncation, each field strength has an equal number of longitudinal and transverse indices (see Table \ref{tab:scalars}) and, consequently, $\sigma$ will not appear in this contribution.

Then the full scalar potential arising from the bulk is
\begin{equation} \label{eq:VBulk}
    V_\text{Bulk}=V_\omega+V_{H_{3}}+\sum_p V_{F_p} \ .
\end{equation}

\subsection{Non-Abelian brane actions}
We want now to understand how the introduction of open string degrees of freedom contributes to the scalar potential of the supergravity theory after compactification. For a stack of $N_{\text{D}6}$ coincident D$6$-branes, the worldvolume action is made up by two contributions, denoted as Dirac-Born-Infeld action and Wess-Zumino action. Following the notation already adopted in \cite{Choi:2018fqw}, \cite{Balaguer:2023jei}, they are
\begin{equation} \label{eq:DBI}
    S^{\textup{DBI}}_{\textrm{D}6}= -T_{\textrm{D}6} \int_{\textup{WV}(\textrm{D}6)} \dd^7 x \; \Tr \bigg( e^{-\hat{\Phi}} \sqrt{-\det(\mathbb{M}_{MN}) \det(\tensor{\mathbb{Q}}{^i_j})} \; \bigg) \quad,
\end{equation}
\begin{equation} \label{eq:WZ}
    S^{\textup{WZ}}_{\textrm{D}6}= \mu_{\textrm{D}6} \int_{\textup{WV}(\textrm{D}6)}  \Tr \bigg\{ \mathrm{P} \bigg[ e^{i \lambda \iota_Y \iota_Y} \Big(\boldsymbol{\hat{C}} \wedge e^{\hat{B}_{(2)}} \Big) \wedge e^{\lambda \mathcal{F}}  \bigg] \bigg\} \quad.
\end{equation}
In these expressions, $\lambda = 2 \pi \ell_s^2$ and we are going to perform expansions for small $\lambda$; this is also the parameter appearing in the brane positions $y^{Ii} = \lambda Y^{Ii}$, with $Y^{Ii}$ living in the gauge algebra of $G_{\textrm{YM}}$ \eqref{eq:GYM}. 

The bosonic effective actions for the orientifold plane, instead, are respectively given by
\begin{equation} \label{eq:OrientifoldDBIAction}
S^{\mathrm{DBI}}_{\mathrm{O}6} = - T_{\mathrm{O}6} \int \mathrm{d}^7x \ e^{- \Phi} \sqrt{- \mathrm{det} (G_{MN})} \quad,
\end{equation}
\begin{equation} \label{eq:WZOrientifold}
S^{\mathrm{WZ}}_{\mathrm{O}6} = \mu_{\mathrm{O}6} \int_{WV(\mathrm{O}6)} \hat{C}_{(7)} + \dots \quad.\end{equation}
Compared to \eqref{eq:DBI} and \eqref{eq:WZ} no other terms appear since at a perturbative level, orientifold planes are not dynamical objects. 
We refer to appendix \ref{appendix:brane_action} for the main details about the conventions we are using.

\subsection{Reduction of the WZ action and modified field strengths}

The Wess-Zumino part of the action does not contribute directly to the scalar potential in $4$ dimensions. Nevertheless, it contains couplings of the R-R fields $C_{(p)}$ that modify the Bianchi identities of the dual fields and consequently the form of the associated field strengths. Such modified field strengths will be the ones appearing in the bulk action ; after compactification, we expect then the modified $F_{(0)}$, $F_{(2)}$, $F_{(4)}$ and $F_{(6)}$ to add new terms to the scalar potential. 

The expression of the potential in 4D gauged supergravity, on the other hand, gives us a strong hint about what these modified field strengths are. Indeed, considering a given $F_{(p)}$, if we select the terms that have the same scaling as $|F_{(p)}|^2$ (this can be done by looking at the scalings in table \ref{Table:Scaling_of_fluxes}) and rearrange all of them in the form of $|\bar{F}_{(p)}|^2$, we can argue that the  $\bar{F}_{(p)}$ defined by this procedure should be precisely the modified field strength we are looking for. However, we want to find explicitly how each of the terms that we expect to be present in the modified field strengths according to the gauged supergravity description can be extracted from compactification of the 10D action, in order to show once again the matching between the two descriptions. This analysis will be, in some points, only qualitative, in the sense that we will often neglect the precise coefficients in front of the terms coming from the WZ action.  

\subsubsection*{Modified $F_{(0)}$}
We start by looking at the modification to the Romans mass $F_{(0)}$. It can be read from a contribution to the WZ action of the form   
\begin{equation}
    \int_{10} C_{(9)} \wedge J_{(1)}^{(\textrm{D}6)} \;,
\end{equation}
or, which is equivalent up to an integration by parts, 
\begin{equation}
    \int_{10} \dd C_{(9)} \wedge \delta {\bar{F}}_{(0)} \;.
\end{equation}
The only possibility to get such a current $J_{(1)}^{(\textrm{D}6)}$ is to start from 
\begin{equation}
\label{WZaction_C9coupling}
     \mu_{\textrm{D}6} \int_{\textup{WV}(\textrm{D}6)}  \Tr \Big\{ \mathrm{P}  \Big[ i \lambda \iota_Y \iota_Y \big(\hat{C}_{(9)} \big)  \Big] \Big\} 
\end{equation}
and rewrite it as  
\begin{equation}
\begin{split}
   & {\mathrm{P} \big[ i \lambda \iota_Y \iota_Y \hat{C}_{(9)}\big]}_{\mu_{0} \mu_{1} \mu_{2} \mu_{3} abc} = \mathrm{P} \big[\lambda \, \tensor{g}{_{JK}^I} Y^{Ji} Y^{Kj} \hat{C}_{\mu_{0} \mu_{1} \mu_{2} \mu_{3} abcij} t_I \big]\\
    & = \mathrm{P} \big[\lambda \, \tensor{g}{_{JK}^I} Y^{Ji} Y^{Kj} \big( C_{\mu_{0} \mu_{1} \mu_{2} \mu_{3} abcij}- \lambda Y^k \big(\tensor{\omega}{_{ik}^d} C_{\mu_{0} \mu_{1} \mu_{2} \mu_{3} abcdj} + \tensor{\omega}{_{ak}^l} C_{\mu_{0} \mu_{1} \mu_{2} \mu_{3} lbcij} \big) \big) t_I \big]\\
    & = \lambda \, \tensor{g}{^I_{JK}} Y^{Ji} Y^{Kj} \big( C_{\mu_{0} \mu_{1} \mu_{2} \mu_{3} abcij}+ \lambda D_a Y^k C_{\mu_{0} \mu_{1} \mu_{2} \mu_{3} kbcij}- \lambda Y^k \tensor{\omega}{_{ak}^l} C_{\mu_{0} \mu_{1} \mu_{2} \mu_{3} lbcij} \big) t_I \;,
\end{split}
\end{equation}
where we have first expanded the hatted field $\hat{C}_{(9)}$, then neglected the term that cancels due to the antisymmetry of $C_{(9)}$ in the indices of type $a,b\ldots$ and applied the pullback to the only term that was not already of order $\lambda^2$. 
The covariant derivative of a $Y$ field with respect to a worldvolume index of type $a,b,\ldots$ becomes (making \eqref{eq:Covariant_Derivative} explicit)
\begin{equation} 
\label{eq:Covariant_Derivative_DaYi}
    D_a Y^i = \partial_a Y^i - i [\mathcal{A}_a, Y^i] = \tensor{\omega}{_{ja}^i} Y^{Ij} t_{I} - \tensor{g}{_{JK}^I} \tensor{\mathcal{A}}{_a^J} Y^{Ki} t_I \; ,
\end{equation}

Then, if we isolate the terms that can be associated to a derivative of the $C_{(9)}$ field, namely those containing at least one metric flux, we get 
\begin{equation}
     \lambda^2 \, \tensor{g}{_{JK}^I} Y^{Ji} Y^{Kj} \big( Y^l \tensor{\omega}{_{la}^k} C_{\mu_{0} \mu_{1} \mu_{2} \mu_{3} kbcij}-  Y^k \tensor{\omega}{_{ak}^l} C_{\mu_{0} \mu_{1} \mu_{2} \mu_{3} lbcij} \big) t_I 
\end{equation}
and take into account the antisymmetry of the indices of the metric flux, we get the expected modification for the field strength $F_{(0)}$. Indeed, we can rewrite this term as 
\begin{equation}
    2 \lambda^2 \, \tensor{g}{_{JK}^I} Y^{Ji} Y^{Kj} Y^{I'l} \tensor{\omega}{_{la}^k} C_{\mu_{0} \mu_{1} \mu_{2} \mu_{3} kbcij} t_I t_{I'} = 2 \lambda^2 \, \tensor{g}{_{JK}^I} Y^{Ji} Y^{Kj} Y^{I'l} t_I t_{I'} \big(\partial_l  C_{\mu_{0} \mu_{1} \mu_{2} \mu_{3} abcij} \big) \;,
\end{equation}
which, in our case with $\mathrm{SO}(3)$ truncation and $\mathfrak{N}=3$, describes a coupling of the exterior derivative $\dd C_{(9)}$ with $g_1 \, Y^3$. This  justifies the result we can get from the gauged supergravity potential:
\begin{equation}
\label{eq:Mod_F0}
    \widetilde{F}_{(0)} = a_3 - g_1 \, Y^3 \; .
\end{equation}

\subsubsection*{Modified $F_{(2)}$}
We move then to considering the modification to the Bianchi identity for $F_{(2)}$. Since we know from type IIA democratic formulation that $F_{(8)}= \star (F_{(2)})$, we want to single out, within the full WZ action, a contribution of the form   
\begin{equation}
    \int_{10} C_{(7)} \wedge J_{(3)}^{(\textrm{D}6)} \;,
\end{equation}
so that the Bianchi identity for the modified field strength, which we denote as $\bar{F}_{(2)}$, becomes
\begin{equation}
    \dd \bar{F}_{(2)} =  J_{(3)}^{(\textrm{D}6)} \;.
\end{equation}
Since the only components of the internal part of $F_{(2)}$ that survive the orientifold projection are $F_{(2)}=\frac{1}{2} F_{ai} e^a \wedge e^i$, the associated Bianchi identity should be of type
\begin{equation}
    \begin{split}
        \dd F_{(2)} =& -\frac{1}{2} \big( F_{mi} \tensor{\omega}{_{an}^m}+ F_{am} \tensor{\omega}{_{in}^m} \big) e^n \wedge e^a \wedge e^i \\
        &= -\frac{1}{2} \big( F_{bi} \tensor{\omega}{_{ac}^b}+ F_{aj} \tensor{\omega}{_{ic}^j} \big) e^c \wedge e^a \wedge e^i \quad .
    \end{split}
\end{equation}
Hence, the components of the current that we should look for are ${(J_{(3)}^{(\textrm{D}6)})}_{abk}$.

\vspace{0.2cm}

The scalar potential suggests that we should expect, after the $\mathrm{SO}(3)$ truncation, a modified field strength 
\begin{equation}
    \tilde{F}_{(2)}=\frac{1}{2} \tilde{F}_{ai} \,  e^a \wedge e^i=\frac{1}{2} \, \tilde{f}_2 \, \delta_{ai} \, e^a \wedge e^i \;,
\end{equation}
where $\tilde{f}_2$ takes the form
\begin{equation}
\label{eq:Mod_f2}
    \tilde{f}_2= a_2 + c_1 Y^2 + g_1 \mathcal{A} Y^2 - a_3 \chi_2 + g_1 Y^3 \chi_2 \;.
\end{equation}
Comparing this expression with the definition of $F_{(2)}$ in the democratic formulation of type IIA supergravity, i.e.
\begin{equation}
     F_{(2)} = \dd C_{(1)} - F_{(0)} B_{(2)} \;,
\end{equation}
we can read off, in the $\mathrm{SO}(3)$ truncation and substituting $F_{(0)}$ with the modified field strength $\bar{F}_{(0)}$ given in \eqref{eq:Mod_F0},
\begin{equation}
    f_2= a_2 - a_3 \chi_2 + g_1 Y^3 \chi_2 \;.
\end{equation}
The remaining terms in order to get the full modified field strength \eqref{eq:Mod_f2} should then necessarily come from the non-Abelian brane action.

The relevant terms in the WZ action are 
\begin{equation}
\label{WZaction_C7coupling}
\begin{split}
     \mu_{\textrm{D}6} \int_{\textup{WV}(\textrm{D}6)}  \Tr \bigg\{ \mathrm{P} & \bigg[ \hat{C}_{(7)} + i \lambda \iota_Y \iota_Y \Big(\hat{C}_{(7)} \wedge \hat{B}_{(2)}\Big)  + i \lambda^2 \iota_Y \iota_Y \hat{C}_{(7)} \wedge \mathcal{F} \\
     & - \frac{\lambda^2}{2} (\iota_Y \iota_Y)^2 \Big(\hat{C}_{(7)} \wedge \hat{B}_{(2)} \wedge \hat{B}_{(2)}\Big) + \mathcal{O}(\lambda^3) \bigg] \bigg\} \quad.
\end{split}
\end{equation}
To evaluate the first term in \eqref{WZaction_C7coupling} we write the pullback of the R-R field as
\begin{equation}
\label{P_C7}
\begin{split}
    \mathrm{P} \big[ \hat{C}_{(7)}\big] _{\mu_{0} \mu_{1} \mu_{2} \mu_{3} abc}= \;& \hat{C}_{\mu_{0} \mu_{1} \mu_{2} \mu_{3} abc} + \lambda D_a Y^i \hat{C}_{\mu_{0} \mu_{1} \mu_{2} \mu_{3} ibc}  \\
    & + \frac{\lambda^2}{2} D_a Y^i D_b Y^j \hat{C}_{\mu_{0} \mu_{1} \mu_{2} \mu_{3} ijc} + \ldots \;,
\end{split}
\end{equation}
where the dots stand for both higher order terms in $\lambda$ and pieces that cannot contribute to the scalar potential. The expansion of the hatted fields around $y^i=0$ gives
\begin{equation}
\begin{split}
     \hat{C}_{\mu_{0} \mu_{1} \mu_{2} \mu_{3} abc} & = C_{\mu_{0} \mu_{1} \mu_{2} \mu_{3} abc} + \lambda Y^i \partial_i C_{\mu_{0} \mu_{1} \mu_{2} \mu_{3} abc}+ \frac{\lambda^2}{2} Y^i Y^j \partial_i \partial_j \hat{C}_{\mu_{0} \mu_{1} \mu_{2} \mu_{3} abc}+ \ldots \\
    & =  C_{\mu_{0} \mu_{1} \mu_{2} \mu_{3} abc} - \lambda Y^i \tensor{\omega}{_{ai}^k} C_{\mu_{0} \mu_{1} \mu_{2} \mu_{3} kbc}+ \frac{\lambda^2}{2} Y^i Y^j \tensor{\omega}{_{ai}^k} \tensor{\omega}{_{bj}^l} \hat{C}_{\mu_{0} \mu_{1} \mu_{2} \mu_{3} klc}+... \;,
\end{split}
\end{equation}
\begin{equation}
\label{C7_ibc}
\begin{split}
     \hat{C}_{\mu_{0} \mu_{1} \mu_{2} \mu_{3} ibc} & = C_{\mu_{0} \mu_{1} \mu_{2} \mu_{3} ibc} + \lambda Y^j \partial_j C_{\mu_{0} \mu_{1} \mu_{2} \mu_{3} ibc}+  \ldots \\
    & =  C_{\mu_{0} \mu_{1} \mu_{2} \mu_{3} ibc} - \lambda Y^j \tensor{\omega}{_{ij}^a} C_{\mu_{0} \mu_{1} \mu_{2} \mu_{3} abc} - \lambda Y^j \tensor{\omega}{_{bj}^k} C_{\mu_{0} \mu_{1} \mu_{2} \mu_{3} ikc}+... \;.
\end{split}
\end{equation}

Using the expression \eqref{eq:Covariant_Derivative_DaYi} for the covariant derivative of the $Y$ fields, the pullback \eqref{P_C7} is left with the following contributions
\begin{equation}
    \begin{split}
        \mathrm{P} & \big[ \hat{C}_{(7)}\big] _{\mu_{0} \mu_{1} \mu_{2} \mu_{3} abc}= \; C_{\mu_{0} \mu_{1} \mu_{2} \mu_{3} abc} - \lambda Y^{Ii} \tensor{\omega}{_{ai}^k} C_{\mu_{0} \mu_{1} \mu_{2} \mu_{3} kbc} t_I\\
        &+ \frac{\lambda^2}{2} \underbrace{Y^{Ii} Y^{I'j} \tensor{\omega}{_{ai}^k} \tensor{\omega}{_{bj}^l} \hat{C}_{\mu_{0} \mu_{1} \mu_{2} \mu_{3} klc}}_{c_1 Y^2} t_I t_{I'}+ \lambda \big( \tensor{\omega}{_{ja}^i} Y^{Ij} - \tensor{g}{_{JK}^I} \tensor{\mathcal{A}}{_a^J} Y^{Ki} \big)  C_{\mu_{0} \mu_{1} \mu_{2} \mu_{3} ibc} t_I\\
        &- \lambda^2 \underbrace{ Y^{I'k}\big( \tensor{\omega}{_{ja}^i} Y^{Ij}- \tensor{g}{_{JK}^I} \tensor{\mathcal{A}}{_a^J} Y^{Ki}}_{c_1 Y^2 \; , \;g_1 \mathcal{A} Y^2}  \big) \big(\tensor{\omega}{_{ik}^a} C_{\mu_{0} \mu_{1} \mu_{2} \mu_{3} abc} + \tensor{\omega}{_{bk}^l} C_{\mu_{0} \mu_{1} \mu_{2} \mu_{3} ilc} \big) t_{I}t_{I'} \\
        & + \frac{\lambda^2}{2} \underbrace{\big( \tensor{\omega}{_{ka}^i} Y^{Ik} - \tensor{g}{_{JK}^I} \tensor{\mathcal{A}}{_a^J} Y^{Ki} \big) \big( \tensor{\omega}{_{lb}^j} Y^{I'l} - \tensor{g}{_{J'K'}^{I'}} \tensor{\mathcal{A}}{_b^{J'}} Y^{K'j} \big)}_{c_1 Y^2 \; , \;g_1 \mathcal{A} Y^2} C_{\mu_{0} \mu_{1} \mu_{2} \mu_{3} ijc} t_I t_{I'}  \;.
    \end{split}
\end{equation}
Here we have written under any of the relevant terms the associated contribution to the modified field strength after we consider the $\mathrm{SO}(3)$ truncation. The terms where only one of the generators $t_I$ appear can always be neglected since they vanish once the trace in \eqref{eq:WZ} is performed.

Let us now consider the second term in \eqref{WZaction_C7coupling}:
\begin{equation}
\begin{split}
    & {i \lambda \iota_Y \iota_Y \big(\hat{C}_{(7)} \wedge \hat{B}_{(2)}\big)}_{\mu_{0} \mu_{1} \mu_{2} \mu_{3} abc} \\
    = &  - i \lambda [Y^i,Y^j] \Big( \hat{C}_{\mu_{0} \mu_{1} \mu_{2} \mu_{3} abc} \hat{B}_{ij}+ \hat{C}_{\mu_{0} \mu_{1} \mu_{2} \mu_{3} ibc} \hat{B}_{aj} + \hat{C}_{\mu_{0} \mu_{1} \mu_{2} \mu_{3} ijc} \hat{B}_{ab}\Big) \\
    = & \,\lambda \, \tensor{g}{_{JK}^I} Y^{Ji} Y^{Kj}  \Big(  \lambda Y^{I'k} \Big( \hat{C}_{\mu_{0} \mu_{1} \mu_{2} \mu_{3} abc} \tensor{H}{_{ijk}}+ \hat{C}_{\mu_{0} \mu_{1} \mu_{2} \mu_{3} ijc} \tensor{H}{_{abk}}\Big) t_{I'} \\
    & + \Big(C_{\mu_{0} \mu_{1} \mu_{2} \mu_{3} ibc} - \lambda Y^{I'k}\tensor{\omega}{_{ik}^d} C_{\mu_{0} \mu_{1} \mu_{2} \mu_{3} dbc} t_{I'} - \lambda Y^{I'k}\tensor{\omega}{_{bk}^l} C_{\mu_{0} \mu_{1} \mu_{2} \mu_{3} ilc} t_{I'} \Big) \hat{B}_{aj} \Big) t_I\;,
\end{split}
\end{equation}
where in the last passage we have substituted the expansion for the hatted field $\hat{C}$ taken from \eqref{C7_ibc} and also expanded the 2-form $\hat{B}_{(2)}$. Since we want to retain only contributions up to order $\lambda^2$, we can neglect any further expansion of the hatted fields, while the only relevant effect of the pullback is in the term 
\begin{multline}
        \mathrm{P}  \big[C_{\mu_{0} \mu_{1} \mu_{2} \mu_{3} ibc} B_{aj}\big] 
         = C_{\mu_{0} \mu_{1} \mu_{2} \mu_{3} ibc} B_{aj} + 2 \lambda D_b Y^{k} C_{\mu_{0} \mu_{1} \mu_{2} \mu_{3} ikc} B_{aj} + \lambda D_a Y^{k} C_{\mu_{0} \mu_{1} \mu_{2} \mu_{3} ibc} B_{kj} \; .
    \end{multline}

We can single out the following contribution to ${(J_{(3)}^{(\textrm{D}6)})}_{abk}$
\begin{equation}
    \lambda^2 \tensor{g}{_{JK}^I} Y^{Ji} Y^{Kj} Y^{ I'k} \tensor{\omega}{_{bk}^l} C_{\mu_{0} \mu_{1} \mu_{2} \mu_{3} ilc} \hat{B}_{aj} t_{I} t_{I'} \;,
\end{equation}
that matches the covariant derivative of the term proportional to $g_1 Y^3 \chi_2$ expected in the modified $\bar{F}_{(2)}$. 

From the third term in \eqref{WZaction_C7coupling} we get :
\begin{equation}
     {\mathrm{P} \big[ i \lambda^2 \iota_Y \iota_Y \hat{C}_{(7)} \wedge \mathcal{F}  \big]}_{\mu_{0} \mu_{1} \mu_{2} \mu_{3} abc} = \,\lambda^2 \, \tensor{g}{_{JK}^I} Y^{Ji} Y^{Kj} C_{\mu_{0} \mu_{1} \mu_{2} \mu_{3} ijc}\tensor{\mathcal{F}}{^{I'}_{ab}} t_I t_{I'} + \mathcal{O}(\lambda^3)\;,
\end{equation}
which cannot be interpreted as bringing a modification to the field strength, while the last one gives no contribution at all since the interior product should be applied to a 11-form.

\subsubsection*{Modified $F_{(4)}$}
In order to find the modified $\bar{F}_{(4)}$, we can start again from the expression of the field strength in the democratic formulation of type IIA \eqref{eq:F4}. However, some modifications are needed. Our setup, indeed, differs from the one of \cite{Bergshoeff:2001pv} in that we are considering the presence of topological fluxes. As explained in detail in \cite{DeWolfe:2005uu}, this implies a modification of the Chern-Simons terms of IIA supergravity, basically because fluxes prevent boundary terms from vanishing when an integration by parts is performed. The general solution for the field strengths ${F}_{(p)}$ in presence of fluxes is then \cite{DallAgata:2009wsi}
\begin{equation} \label{eq:F_DallAgata}
    \boldsymbol{F}  = \dd \boldsymbol{C}+ H_{(3)} \wedge \boldsymbol{C}  + \boldsymbol{F}^{\textup{flux}} \wedge e^{-B_{(2)}} \;,
\end{equation}
where bold symbols $\boldsymbol{F}$, $\boldsymbol{C}$, $\boldsymbol{F}^{\textup{flux}}$ denote respectively formal sums of all the field strengths, the RR fields and the constant flux contributions to the field strengths (for any field strength ${F}_{(p)}$ on the LHS we have to consider only the terms with the appropriate degree in the above expansion on the RHS).

In this case, this means that $F_{(4)}$ (before the contributions coming from the open string sector) can be written as
\begin{equation}
    F_{(4)} = \dd C_{(3)} - F_{(2)}^{\textup{flux}} \wedge B_{(2)} +\frac{1}{2} F_{(0)}^{\textup{flux}} B_{(2)} \wedge B_{(2)} + F_{(4)}^{\textup{flux}} \;.
\end{equation}
In our setup, namely the orientifold projection and  the $\mathrm{SO}(3)$ truncation, the only non-vanishing components of the field strength are 
\begin{equation}
\label{NonModF4_components}
    F_{(4)}=\frac{1}{4!} F_{aibj} \,  e^a \wedge e^i \wedge e^b \wedge e^j=\frac{1}{4!} \, f_4 \, \varepsilon_{abc} \, \varepsilon_{ijk}  \, \delta^{ck} \, e^a \wedge e^i \wedge e^b \wedge e^j \;.
\end{equation}
If we consider the following expressions  
\begin{equation}
    \big( \dd C_{(3)} \big)_{aibj}= - \tensor{\omega}{_{ab}^c} C_{cij} - \tensor{\omega}{_{ij}^c} C_{cab} + 2 \tensor{\omega}{_{aj}^k} C_{kib} \;,
\end{equation}
\begin{equation}
    \big(  F_{(2)} \wedge B_{(2)} -\frac{1}{2} F_{(0)} B_{(2)} \wedge B_{(2)}\big)_{aibj}= 2 F_{ai} B_{bj} - F_{(0)} B_{ai} B_{bj} \;,
\end{equation}
and compare them with \eqref{NonModF4_components}, taking into account also the fact that we have already obtained the full expression for the modified field strengths $\bar{F}_{(2)}$ and $\bar{F}_{(0)}$, we obtain 
\begin{equation}
    f_4=  \,a_1  + 
 \bar{c}_1 \chi_1 - b_1 \chi - 2 c_1 \chi_1 - 2 (a_2 + c_1 Y^2 + g_1 \mathcal{A} Y^2 - a_3 \chi_2 + g_1 Y^3 \chi_2 ) \chi_2 - ( a_3 - g_1 Y^3) \chi_2^2 \;.
\end{equation}
On the other hand, the modified field strength $\tilde{F}_{(4)}$ that we expect from the scalar potential in the 4-dimensional gauged supergravity theory suggests that the scalar $f_4$ is replaced by 
\begin{equation}
\label{eq:Mod_f4}
\begin{split}
    \tilde{f}_4= & \,a_1 - b_1 \chi - 2 c_1 \chi_1 + 
 \bar{c}_1 \chi_1 - 2 a_2 \chi_2 - 2 c_1 Y^2 \chi_2 - 2 g_1 \mathcal{A} Y^2 \chi_2  \\
 & + a_3 \chi_2^2 -  g_1 Y^3 \chi_2^2 - g_0 Y - g_1 \mathcal{A}^2 Y - \frac{c_0 Y^2}{2} - \frac{1}{2} \mathcal{A} Y (2 c_1 + \bar{c}_1) \;.
\end{split}
\end{equation}
The missing terms in $f_4$, namely
\begin{equation}
    \Delta f_4 = \tilde{f}_4 -f_4 = - g_0 Y - g_1 \mathcal{A}^2 Y - \frac{c_0 Y^2}{2} - \frac{1}{2} \mathcal{A} Y (2 c_1 + \bar{c}_1) \;,
\end{equation}
should then come out from the WZ action. In order to find them, or equivalently to find the modified Bianchi identity for the field strength $F_{(4)}$, we have to look at the couplings in the Wess-Zumino action that involve the ${C}_{(5)}$ field:
\begin{equation}
\label{WZaction_C5coupling}
\begin{split}
     \mu_{\textrm{D}6} \int_{\textup{WV}(\textrm{D}6)}  \Tr \bigg\{ \mathrm{P} & \bigg[ \hat{C}_{(5)} \wedge \hat{B}_{(2)} + \frac{i \lambda}{2} \iota_Y \iota_Y \Big(\hat{C}_{(5)} \wedge \hat{B}_{(2)}\wedge \hat{B}_{(2)}\Big)  +  \lambda \, \hat{C}_{(5)} \wedge \mathcal{F} \\
     & + i \lambda^2 \iota_Y \iota_Y  \Big(\hat{C}_{(5)} \wedge \hat{B}_{(2)} \Big)  \wedge \mathcal{F}+ \mathcal{O}(\lambda^3) \bigg] \bigg\} \quad,
\end{split}
\end{equation}
aiming at isolating a current $J_{(5)}^{(\textrm{D}6)}$ appearing in
\begin{equation}
    \int_{10} C_{(5)} \wedge J_{(5)}^{(\textrm{D}6)} \;.
\end{equation} 

The fluxes obtained from dimensional reduction of $F_{(4)}$ that survive the orientifold projection are $F_{(4)}=\frac{1}{4!} F_{aibj} e^a \wedge e^i \wedge e^b \wedge e^j$, then \begin{equation}
    \begin{split}
        \dd F_{(4)} =& -\frac{1}{4!} \big( F_{mibj} \tensor{\omega}{_{an}^m}+ F_{ambj} \tensor{\omega}{_{in}^m}+F_{aimj} \tensor{\omega}{_{bn}^m}+ F_{aibm} \tensor{\omega}{_{jn}^m} \big) e^n \wedge e^a \wedge e^i\wedge e^b \wedge e^j \\
        =& -\frac{1}{4!} \big( F_{cibj} \tensor{\omega}{_{ad}^c}+ F_{akbj} \tensor{\omega}{_{id}^k}+F_{aicj} \tensor{\omega}{_{bd}^c}+ F_{aibk} \tensor{\omega}{_{jd}^k} \big) e^d \wedge e^a \wedge e^i\wedge e^b \wedge e^j \; .
    \end{split}
\end{equation}
Hence, we look for components of the current of type ${(J_{(5)}^{(\textrm{D}6)})}_{abcij}$, which should be coupled with a ${(C_{(5)})}_{\mu_{0} \mu_{1} \mu_{2} \mu_{3} k}$ field.

The first term in \eqref{WZaction_C5coupling} can be rewritten as
\begin{equation}
\label{P_C5B2}
\begin{split}
    \mathrm{P}& \big[ \hat{C}_{(5)} \wedge \hat{B}_{(2)}\big] _{\mu_{0} \mu_{1} \mu_{2} \mu_{3} abc}= \hat{C}_{\mu_{0} \mu_{1} \mu_{2} \mu_{3} a} \hat{B}_{bc}+ \lambda D_a Y^i \hat{C}_{\mu_{0} \mu_{1} \mu_{2} \mu_{3} i} \hat{B}_{bc}+ \lambda D_b Y^i \hat{C}_{\mu_{0} \mu_{1} \mu_{2} \mu_{3} a} \hat{B}_{ic}\\
    & + \lambda^2 D_a Y^i D_b Y^j \hat{C}_{\mu_{0} \mu_{1} \mu_{2} \mu_{3} i}  \hat{B}_{jc}+  \lambda^2 D_a Y^i D_b Y^j \hat{C}_{\mu_{0} \mu_{1} \mu_{2} \mu_{3} c}  \hat{B}_{ij}+ \ldots \\
    = &\,  \lambda Y^{i} {C}_{\mu_{0} \mu_{1} \mu_{2} \mu_{3} a} H_{bci} - \underbrace{\lambda^2 Y^{i} Y^{j} \tensor{\omega}{_{ai}^k}{C}_{\mu_{0} \mu_{1} \mu_{2} \mu_{3} k}  H_{bcj} }_{c_0 Y^2}- \underbrace{\lambda^2 \big( D_a Y^i \big) Y^{j} {C}_{\mu_{0} \mu_{1} \mu_{2} \mu_{3} i} H_{bcj}}_{c_0 Y^2}\\
    &+ \lambda D_b Y^i {C}_{\mu_{0} \mu_{1} \mu_{2} \mu_{3} a} B_{ic}\underbrace{-\lambda^2 \big( D_b Y^i \big) Y^{j} \tensor{\omega}{_{aj}^k}{C}_{\mu_{0} \mu_{1} \mu_{2} \mu_{3} k} B_{ic}}_{g_1 A Y^2 \chi_2}+ \underbrace{\lambda^2 D_a Y^i D_b Y^j {C}_{\mu_{0} \mu_{1} \mu_{2} \mu_{3} i}  B_{jc}}_{g_1 A Y^2 \chi_2 \;, \;c_1 Y^2 \chi_2} \;.
\end{split}
\end{equation}
With the same procedure, we get from the second contribution to \eqref{WZaction_C5coupling}
\begin{equation}
\begin{split}
   & {\mathrm{P} \big[i \lambda \iota_Y \iota_Y \big(\hat{C}_{(5)} \wedge \hat{B}_{(2)} \wedge \hat{B}_{(2)}\big) \big]}_{\mu_{0} \mu_{1} \mu_{2} \mu_{3} abc} \\
    = & \mathrm{P} \Big[ - i \lambda [Y^i,Y^j] \Big( \hat{C}_{\mu_{0} \mu_{1} \mu_{2} \mu_{3} a} \hat{B}_{bc} \hat{B}_{ij}+ \hat{C}_{\mu_{0} \mu_{1} \mu_{2} \mu_{3} a} \hat{B}_{bi} \hat{B}_{jc} + \hat{C}_{\mu_{0} \mu_{1} \mu_{2} \mu_{3} i} \hat{B}_{ja} \hat{B}_{bc} \Big) \Big] \\
    = & \mathrm{P} \Big[ - i \lambda [Y^i,Y^j] \Big( C_{\mu_{0} \mu_{1} \mu_{2} \mu_{3} a} B_{bi} B_{jc} - \lambda Y^k \tensor{\omega}{_{ak}^l} C_{\mu_{0} \mu_{1} \mu_{2} \mu_{3} l} B_{bi} B_{jc} \\
    & \quad + \lambda Y^k C_{\mu_{0} \mu_{1} \mu_{2} \mu_{3} i} B_{ja} H_{bck} \Big) \Big] \\
    = & \,\lambda \, \tensor{g}{_{JK}^I} Y^{Ji} Y^{Kj}  \Big( C_{\mu_{0} \mu_{1} \mu_{2} \mu_{3} a} B_{bi} B_{jc} + \underbrace{\lambda  D_a Y^k C_{\mu_{0} \mu_{1} \mu_{2} \mu_{3} k} B_{bi} B_{jc}}_{g_1 Y^3 \chi_2^2} \\
    & + \lambda^2 Y^{I'k} \Big(- \underbrace{ \tensor{\omega}{_{ak}^l} C_{\mu_{0} \mu_{1} \mu_{2} \mu_{3} l} B_{bi} B_{jc}}_{g_1 Y^3 \chi_2^2} + C_{\mu_{0} \mu_{1} \mu_{2} \mu_{3} i} B_{ja} H_{bck} \Big) t_{I'} \Big) t_I\;.
\end{split}
\end{equation}
From the third term in \eqref{WZaction_C5coupling} we can obtain
\begin{equation}
\begin{split}
    & {\mathrm{P} \big[\lambda \, \hat{C}_{(5)} \wedge \mathcal{F} \big]}_{\mu_{0} \mu_{1} \mu_{2} \mu_{3} abc} \\
    = &\;  \lambda  \hat{C}_{\mu_{0} \mu_{1} \mu_{2} \mu_{3} a} \mathcal{F}_{bc}  + \lambda^2 D_a Y^i \hat{C}_{\mu_{0} \mu_{1} \mu_{2} \mu_{3} i} \mathcal{F}_{bc}\\
    = &\;  \lambda  C_{\mu_{0} \mu_{1} \mu_{2} \mu_{3} a} \mathcal{F}_{bc}  -\lambda^2 Y^i \tensor{\omega}{_{ai}^j} C_{\mu_{0} \mu_{1} \mu_{2} \mu_{3} j} \mathcal{F}_{bc} + \lambda^2 D_a Y^i C_{\mu_{0} \mu_{1} \mu_{2} \mu_{3} i} \mathcal{F}_{bc} \\
    = &\; \lambda  C_{\mu_{0} \mu_{1} \mu_{2} \mu_{3} a} \mathcal{F}_{bc}  - 2 \lambda^2 Y^i \tensor{\omega}{_{ai}^j} C_{\mu_{0} \mu_{1} \mu_{2} \mu_{3} j} \mathcal{F}_{bc} - \lambda^2 \tensor{g}{_{JK}^I} \mathcal{A}^J_a Y^{Ki} t_I C_{\mu_{0} \mu_{1} \mu_{2} \mu_{3} i} \mathcal{F}_{bc} \;.
\end{split}
\end{equation}
In components, the full expression of the non-abelian field strength  \eqref{eq:Nonab_FieldStrength} is
\begin{equation}
\label{eq:Nonab_FieldStrength_components}
    \mathcal{F}_{ab} = \tensor{\mathcal{F}}{^I_{ab}} t_{I} = \Big( {\tensor{\mathcal{F}}{^I_{ab}}}^{\textup{(flux)}} + \tensor{g}{_{JK}^I} \mathcal{A}^J_a \mathcal{A}^K_b + \tensor{\omega}{_{[ab]}^c} \mathcal{A}^I_c \Big)t_{I} \;.
\end{equation}
Using this result, we can isolate in the previous formula the contributions containing at least a metric flux, and then contributing to our modified Bianchi identity:
\begin{equation}
\begin{split}
    & -2 \lambda^2 \tensor{\omega}{_{ai}^j} C_{\mu_{0} \mu_{1} \mu_{2} \mu_{3} j} \Big(\underbrace{Y^{Ii} {\tensor{\mathcal{F}}{^{I'}_{bc}}}^{\textup{(flux)}} }_{g_0 Y} + \underbrace{Y^{Ii} \tensor{g}{_{J'K'}^{I'}} \mathcal{A}^{J'}_b \mathcal{A}^{K'}_c}_{g_1 Y \mathcal{A}^2} \Big) t_{I} t_{I'}  \\
    & -2 \lambda^2 C_{\mu_{0} \mu_{1} \mu_{2} \mu_{3} j} \underbrace{\tensor{\omega}{_{ai}^j} \tensor{\omega}{_{bc}^d}Y^{Ii}  \mathcal{A}^{I'}_d }_{\bar{c}_1 Y \mathcal{A} \;, \;c_1 Y \mathcal{A}} t_{I} t_{I'} + \lambda^2  C_{\mu_{0} \mu_{1} \mu_{2} \mu_{3} i} \tensor{\omega}{_{bc}^d} \underbrace{\tensor{g}{_{JK}^I} \mathcal{A}^J_a Y^{Ki}  \mathcal{A}^{I'}_d }_{g_1 Y \mathcal{A}^2} t_{I} t_{I'} \;.
\end{split}
\end{equation}

The last term in \eqref{WZaction_C5coupling} is already of order $\lambda^2$, then we can neglect both the expansion of the fields and the pullback:
\begin{equation}
\begin{split}
    & {\mathrm{P} \big[ i \lambda^2 \iota_Y \iota_Y  \Big(\hat{C}_{(5)} \wedge \hat{B}_{(2)} \Big)  \wedge \mathcal{F} \big]}_{\mu_{0} \mu_{1} \mu_{2} \mu_{3} abc}  \\
    = & \,\lambda^2 \, \tensor{g}{^I_{JK}} Y^{Ji} Y^{Kj} C_{\mu_{0} \mu_{1} \mu_{2} \mu_{3} i} B_{ja} \mathcal{F}_{bc}  t_I  + \mathcal{O}(\lambda^3) \;,
\end{split}
\end{equation} 
then it is expected that it gives no contribution to the modified Bianchi identity.

\subsubsection*{Modified $F_{(6)}$}
According to the formula \eqref{eq:F_DallAgata} for the R-R field strengths of type IIA supergravity, the full expression for $F_{(6)}$ is
\begin{equation}
    F_{(6)} = \dd C_{(5)} + H_{(3)} \wedge C_{(3)} - F_{(4)}^{\textup{flux}} \wedge B_{(2)} +\frac{1}{2} F_{(2)}^{\textup{flux}} \wedge B_{(2)} \wedge B_{(2)} -\frac{1}{3!} F_{(0)}^{\textup{flux}} B_{(2)} \wedge B_{(2)} \wedge B_{(2)} + F_{(6)}^{\textup{flux}} \;.
\end{equation}

In the $\mathrm{SO}(3)$ truncation, the field strength has the form 
\begin{equation}
    F_{(6)}=\frac{1}{6!} F_{aibjck} \,  e^a \wedge e^i \wedge e^b \wedge e^j \wedge e^c \wedge e^k=\frac{1}{6!} \, f_6 \, \varepsilon_{abc} \, \varepsilon_{ijk} \, e^a \wedge e^i \wedge e^b \wedge e^j \wedge e^c \wedge e^k \;.
\end{equation}
In order to derive the expression for the scalar $f_6$, we can write in components
\begin{equation}
    \big( H_{(3)} \wedge C_{(3)}\big)_{aibjck} = H_{ijk} C_{abc} + 3 H_{abk} C_{ijc}
\end{equation}
\begin{equation}
\begin{split}
    & \big(  F_{(4)} \wedge B_{(2)} -\frac{1}{2} F_{(2)} \wedge B_{(2)} \wedge B_{(2)} +\frac{1}{3!} F_{(0)} B_{(2)} \wedge B_{(2)} \wedge B_{(2)}\big)_{aibjck} \\
    =& \, 3\, F_{aibj} B_{ck} -\frac{6}{2} F_{ai} B_{bj} B_{ck}  +\frac{6}{3!} F_{(0)} B_{ai} B_{bj} B_{ck} \;,
\end{split}    
\end{equation}
so that we get 
\begin{equation}
    f_6=  \,a_0  - b_0 \chi +  3 c_0 \chi_1 -3 \chi_2 \tilde{f}_4 -3 \chi_2^2 \tilde{f}_2 - \chi_2^3 \tilde{F}_{(0)} \;,
\end{equation}
where we have to consider the previous results for the modified field strengths of rank $p<6$. After these replacements, we can compare $f_6$ with the modified field strength computed from the gauged supergravity potential
\begin{equation}
\label{eq:Mod_f6}
\begin{split}
    \tilde{f}_6=&  \, a_0- b_0 \chi + 3 c_0 \chi_1 - 3 a_1 \chi_2 + 
 3 g_0 Y \chi_2 + \frac{3}{2} c_0 Y^2 \chi_2 + 3 b_1 \chi \chi_2 +  6 c_1 \chi_1 \chi_2 - 3 \bar{c}_1 \chi_1 \chi_2 \\
 & +3 a_2 \chi_2^2 +  3 c_1 Y^2 \chi_2^2 - a_3 \chi_2^3 + g_1 Y^3 \chi_2^3 + 3 \mathcal{A} g_0 + g_1 \mathcal{A}^3  +  \frac{3}{2} \mathcal{A}^2 (\bar{c}_1 + 2 g_1 Y \chi_2)  \\
 & + \frac{3}{2} \mathcal{A} Y (c_0 + \chi_2 (2 c_1 + \bar{c}_1 + 2 g_1 Y \chi_2))\;.
\end{split}
\end{equation}
Hence, the terms that we should be able to read off from the non-abelian brane action correspond to
\begin{equation}
    \Delta f_6 = \tilde{f}_6 -f_6 = \frac{3}{2} \bar{c}_1 \mathcal{A}^2 + 3 \mathcal{A} g_0 + \mathcal{A}^3 g_1 + \frac{3}{2} c_0 \mathcal{A} Y \;.
\end{equation}

The contributions to the WZ action that we can analyze concerning $\tilde{F}_{(6)}$ are 
\begin{equation}
\label{WZaction_C3coupling}
\begin{split}
     \mu_{\textrm{D}6} \int_{\textup{WV}(\textrm{D}6)}  \Tr \bigg\{ \mathrm{P} & \bigg[ \frac{1}{2} \hat{C}_{(3)} \wedge \hat{B}_{(2)} \wedge \hat{B}_{(2)} + \frac{i \lambda}{3 !} \iota_Y \iota_Y \Big(\hat{C}_{(3)} \wedge \hat{B}_{(2)} \wedge \hat{B}_{(2)} \wedge \hat{B}_{(2)}\Big)  \\ 
     & +\lambda \, \hat{C}_{(3)} \wedge \hat{B}_{(2)} \wedge \mathcal{F}
     + i \lambda^2 \iota_Y \iota_Y  \Big(\hat{C}_{(3)} \wedge \hat{B}_{(2)}  \wedge \hat{B}_{(2)} \Big)  \wedge \mathcal{F} \\
     & + \frac{\lambda^2}{2} \hat{C}_{(3)} \wedge \mathcal{F} \wedge \mathcal{F} + \mathcal{O}(\lambda^3) \bigg] \bigg\} \quad.
\end{split}
\end{equation}
Let us focus on the last contribution. Writing it in components, we can select a term of the form
\begin{equation}
   \Big( \hat{C}_{(3)} \wedge \mathcal{F} \wedge \mathcal{F} \Big)_{\mu_{0} \mu_{1} \mu_{2} \mu_{3} abc} = \hat{C}_{\mu_{0} \mu_{1} \mu_{2}} \mathcal{F}_{\mu_3 a} \mathcal{F}_{bc} = \hat{C}_{\mu_{0} \mu_{1} \mu_{2}} \big( \partial_{\mu_3} \mathcal{A}_{a} \big) \mathcal{F}_{bc} + \ldots \;.
\end{equation}
After an integration by parts, it gives back in the WZ action a contribution $\partial_{\mu_3} \hat{C}_{\mu_{0} \mu_{1} \mu_{2}} \mathcal{A}_{a} \mathcal{F}_{bc}$, that we can interpret as a coupling of the derivative of ${C}_{(3)}$ with the modified field strength $\tilde{F}_{(6)}$. Using the expression \eqref{eq:Nonab_FieldStrength_components} for the non-abelian field strength, we get
\begin{equation}
    \partial_{\mu_3} \hat{C}_{\mu_{0} \mu_{1} \mu_{2}}   \Big(\underbrace{\mathcal{A}_{a}^I {\tensor{\mathcal{F}}{^{I'}_{bc}}}^{\textup{(flux)}} }_{g_0 \mathcal{A}} - \underbrace{\mathcal{A}_{a}^I \tensor{g}{_{J'K'}^{I'}} \mathcal{A}^{J'}_b \mathcal{A}^{K'}_c}_{g_1 \mathcal{A}^3} + \underbrace{\mathcal{A}_{a}^I \tensor{\omega}{_{bc}^d} \mathcal{A}_d^{I'}}_{\bar{c}_1 \mathcal{A}^2}\Big) t_{I} t_{I'} \;.
\end{equation}
With a similar procedure, starting from the third term in \eqref{WZaction_C3coupling}, we can derive from
\begin{equation}
   \Big( \hat{C}_{(3)} \wedge \hat{B}_{(2)} \wedge \mathcal{F} \Big)_{\mu_{0} \mu_{1} \mu_{2} \mu_{3} abc} = \hat{C}_{\mu_{0} \mu_{1} \mu_{2}} \mathcal{F}_{\mu_3 a} \hat{B}_{bc} = \hat{C}_{\mu_{0} \mu_{1} \mu_{2}} \big( \partial_{\mu_3} \mathcal{A}_{a} \big) \hat{B}_{bc} + \ldots 
\end{equation}
a coupling $\partial_{\mu_3} \hat{C}_{\mu_{0} \mu_{1} \mu_{2}} \mathcal{A}_{a} \hat{B}_{bc}$ that finally gives, if we replace the expansion for the hatted field $\hat{B}_{(2)}$,
\begin{equation}
    \partial_{\mu_3} \hat{C}_{\mu_{0} \mu_{1} \mu_{2}} \lambda    \Big(\underbrace{\mathcal{A}_{a}^I Y^{I'i} H_{bci} }_{c_0 \mathcal{A} Y} \Big) t_{I} t_{I'} \;.
\end{equation}

\subsection{Reduction of the DBI action}
The DBI action \eqref{eq:DBI} is responsible for contributions to the scalar potential in 4 dimensions. We expand the metric and the B-field around the position $y^{Ii} = \lambda Y^{Ii}$ (static gauge), at first order in $\lambda$. Furthermore, we only keep the fluxes with non-contracted legs in the worldvolume directions of the D6-brane. We obtain:
\begin{align} \label{eq:GB}
    \hat{G}_{ab} = \rho \sigma^2 \delta_{ab} \quad, 
    \quad \hat{G}_{ai} & = 0 \qquad \quad, \quad  \hat{G}_{ij} = \rho \sigma^{-2} \delta_{ij} \quad, \\
    \hat{B}_{ab} = \lambda c_0 \ \varepsilon_{abk} Y^k \quad, \quad
    \hat{B}_{ai} & = \chi_2 \delta_{ai} \quad, \quad
    \hat{B}_{ij} = 0 \qquad \qquad.
    \label{eq:GB1}
\end{align}
According to the definition \eqref{eq:EMN}, one can compute the fields 
\begin{equation} \label{eq:Ehat}
\hat{E}_{ab} = \rho \sigma^2 \delta_{ab} + \lambda c_0 \varepsilon_{abk} Y^k \quad, \quad \hat{E}_{ai} = - \hat{E}_{ia} = \chi_2 \delta_{ai} \quad, \quad \hat{E}_{ij} = \rho \sigma^{-2} \delta_{ij} \quad .
\end{equation}
Due to the definition \eqref{eq:Qij}, we obtain that 
\begin{equation*}
    \tensor{\mathbb{Q}}{^i_j}=\tensor{\delta}{^i_j} + \lambda \tensor{A}{^i_j} \quad,
\end{equation*}
with 
\[
\tensor{A}{^i_j} = \rho \sigma^{-2} g_1 {\varepsilon_{IJ}}^L Y^{Ii} \tensor{Y}{^J_j} t_L \quad.
\]
The antisymmetry property of ${\varepsilon_{IJ}}^L$ leads to $\tensor{A}{^i_i} = 0$, hence 
\begin{equation} \label{eq:sqrtQ}
\sqrt{\det \tensor{\mathbb{Q}}{^i_j}} = \ 1 - \frac{\lambda^2}{2} \Bigl (\frac{{A^i}_j {A^j}_i}{2} \Bigr) + \dots = 1 + \frac{\lambda^2}{2} \rho^2 \sigma^{-4} {g_1}^2 {Y}^2 \sum_{I=1}^\mathfrak{N} {t_I}^2 + \dots \quad.
\end{equation}
In the last step, the definitions \eqref{eq:openaxions} and \eqref{eq:openstring} have been used. 

In order to find the contribution of \eqref{eq:MMN} to the DBI action, let us employ \eqref{eq:PullBack} to compute 
\begin{equation} \label{eq:pullbackEab}
P [\hat{E}_{ab}] = \hat{E}_{ab} + \lambda D_a Y^k \hat{E}_{kb} + \lambda D_b Y^k \hat{E}_{ak} + \frac{\lambda^2}{2} D_a Y^i D_b Y^j \hat{E}_{ij} \quad.
\end{equation}
In particular, by using the expressions \eqref{eq:Ehat} and the definition \eqref{eq:Covariant_Derivative}, one obtains
\begin{align*}
D_a Y^k \hat{E}_{kb} & = - \chi_2 \delta_{kb} \bigl ( c_1 {\varepsilon_{am}}^k Y^m - g_1 {\varepsilon_{IJ}}^K {\mathcal{A}_a}^I Y^{Jk} t_K \bigr) \quad, \\
D_b Y^k \hat{E}_{ak} & = \chi_2 \delta_{ak} \bigl ( c_1 {\varepsilon_{bm}}^k Y^m - g_1 {\varepsilon_{IJ}}^K {\mathcal{A}_b}^I Y^{Jk} t_K \bigr) \quad, \\
D_a Y^i D_b Y^j \hat{E}_{ij} & = \rho \sigma^{-2} \bigl ( c_1 {\varepsilon_{am}}^i Y^m - g_1 {\varepsilon_{IJ}}^K {\mathcal{A}_a}^I Y^{Ji} t_K \bigr) \bigl ( c_1 {\varepsilon_{bl}}^j Y^l - g_1 {\varepsilon_{MN}}^L {\mathcal{A}_b}^M Y^{Nj} t_L \bigr) \delta_{ij} \quad.
\end{align*}
Accordingly, \eqref{eq:pullbackEab} can be recast in 
\begin{equation} \label{eq:pullbackEab1}
P[\hat{E}_{ab}] = \rho \sigma^2 \delta_{ab} + \lambda \hat{E}^{(1)}_{ab} + \lambda^2 \hat{E}^{(2)}_{ab} \quad,
\end{equation}
with 
\begin{align*}
\hat{E}^{(1)}_{ab} = c_0 \varepsilon_{abk} Y^k + & 2 \chi_2 \bigl (c_1 \varepsilon_{abm} Y^m - g_1 {\varepsilon_{IJ}}^K {\mathcal{A}_{[b|}}^I {Y^J}_{|a]} t_K \bigr ) \quad, \\
& \hat{E}^{(2)}_{ab} = \frac{1}{2} D_a Y^i D_b Y^j \hat{E}_{ij} \quad.
\end{align*}
Moreover, 
\[
{({\mathbb{Q}}^{-1})^i}_j = \tensor{\delta}{^i_j} - \lambda \tensor{A}{^i_j} + \lambda^2 ( \tensor{A}{^i_l}\tensor{A}{^l_j}) + \dots \quad.
\]
This enters the matrix $\mathbb{M}_{MN}$ defined in \eqref{eq:MMN}, along with 
\begin{align*}
P[\hat{E}_{ai}] & = \hat{E}_{ai} + \lambda D_a Y^k \hat{E}_{ki} = \chi_2 \delta_{ai} + \lambda \hat{E}^{(1)}_{ai} \quad, \\ 
P[\hat{E}_{jb}] & = \hat{E}_{jb} + \lambda D_b Y^k \hat{E}_{jk} = - \chi_2 \delta_{jb} + \lambda \hat{E}^{(1)}_{bj} \quad.
\end{align*}
where we make use of
\begin{equation}
    \hat{E}^{(1)}_{ai} = \rho \sigma^{-2} \bigl (c_1 \varepsilon_{iam} Y^m - g_1 {\varepsilon_{IJ}}^L {\mathcal{A}_a}^I Y^{Ji} t_L \bigr) \quad.
\end{equation}
We can then calculate 
\begin{equation*}
P[\hat{E}_{ai} ({\mathbb{Q}}^{-1} - \delta)^{ij} \hat{E}_{jb}] = \rho \sigma^{-2} \bigl [ \lambda {\chi_2}^2 \delta_{ai}A^{ij} \delta_{jb} + \lambda^2 \bigl (2 \chi_2 \hat{E}^{(1)}_{[a|i} A^{ij} \delta_{j|b]} - {\chi_2}^2 {A^i}_l A^{lj} \delta_{ai} \delta_{jb}  \bigr ) \bigr] + \dots
\end{equation*}
By also writing $\mathcal{F}_{ab}$ as in \eqref{eq:Nonab_FieldStrength_components}
\begin{equation}
\mathcal{F}_{ab} = \bigl ( g_0 + {\bar{c}_1} \mathcal{A} + g_1 \mathcal{A}^2 \bigr ) {\varepsilon_{ab}}^I t_I \quad,
\end{equation}
one can compute the matrix $\mathbb{M}_{MN}$. The relevant terms for the potential, up to second order in $\lambda$, will be
\begin{align*}
\mathbb{M}_{\mu \nu} & = \tau^{-2} g_{\mu \nu} + \dots \quad, \\
\mathbb{M}_{ab} & = \rho \sigma^2 \delta_{ab} + \lambda {\mathbb{M}^{(1)}}_{ab} + \lambda^2 {\mathbb{M}^{(2)}}_{ab} + \dots \quad,
\end{align*}
with 
\begin{align*}
    {\mathbb{M}^{(1)}}_{ab} & = \hat{E}^{(1)}_{ab} + \rho \sigma^{-2} {\chi_2}^2 \delta_{ai} A^{ij} \delta_{jb} + \mathcal{F}_{ab} \quad, \\
    {\mathbb{M}^{(2)}}_{ab} & = \hat{E}^{(2)}_{ab} + 2 \chi_2 \hat{E}^{(1)}_{[a|i} A^{ij} \delta_{j|b]} - {\chi_2}^2 {A^i}_l A^{lj} \delta_{ai} \delta_{jb} \quad.
\end{align*}
Furthermore, since ${\mathbb{M}^{(1)}}_{ab}$ is antisymmetric in $(a, b)$, 
\begin{equation} \label{eq:sqrtM}
    \sqrt{\det \mathbb{M}_{MN}} = \sqrt{-g^{(4)}} \tau^{-4} \rho^{3/2} \sigma^3 \Bigl (1 + \frac{\lambda^2}{2} \rho^{-1} \sigma^{-2} {{\mathbb{M}^{(2)}}^a}_{a} - \frac{\lambda^2}{4} \rho^{-2} \sigma^{-4} {{\mathbb{M}^{(1)}}^a}_{b} {{\mathbb{M}^{(1)}}^b}_{a} \Bigr ) + \dots \quad.
\end{equation}
The calculation of the DBI action for each D6-brane finally requires to insert \eqref{eq:dilatonscaling}, \eqref{eq:sqrtQ} and \eqref{eq:sqrtM} in \eqref{eq:DBI} and to exploit the condition \eqref{eq:GeneratorRelations} on the gauge generators 
\begin{equation} 
    [t_I, t_J] = -i {g_{IJ}}^K t_K \quad, \quad \mathrm{Tr} [t_I, t_J] = \delta_{IJ} \quad.
\end{equation}
The setup under study also contains an orientifold plane $\mathrm{O}6^-$, whose DBI and WZ actions are given by \eqref{eq:OrientifoldDBIAction} and \eqref{eq:WZOrientifold}. The contribution of O6-planes in the DBI reduction only appears at the zeroth order in $\lambda$ and can be obtained by using \eqref{eq:compactmetric} and \eqref{eq:dilatonscaling} in \eqref{eq:OrientifoldDBIAction}. Both considering the latter and the contribution of the D6-branes \eqref{eq:DBI}, the DBI potential reads
\begin{equation} \label{eq:VDBIsecond}
    V_{\mathrm{DBI}} = V^{(0)}_{\mathrm{DBI}} + \lambda^2 V^{(2)}_{\mathrm{DBI}} + \dots \quad.
\end{equation}
In particular, both D6-branes and O6-orientifold planes give rise to
\begin{equation}
V_{\mathrm{DBI}}^{(0)} = \sigma^3 \tau^{-3} \bigl (N_{D6} T_{D6} + T_{\mathrm{O}6} \bigr ) \quad.
\end{equation}
This can be rewritten on grounds of the tadpole cancellation condition. Indeed, the modified Bianchi identity \eqref{eq:modifiedBianchiF2} in presence of D6/O6 sources reads 
\begin{equation} \label{eq:modifiedtadpole}
\mathrm{d} F_{(2)} - H_{(3)} \wedge F_{(0)} = Q_6 \mathrm{vol}_{\mathcal{M}_3^{(a)}} \quad,
\end{equation}
with on the right hand side an effective current density $Q_6$ and the volume of the submanifold of the three compact directions along the worldvolume of the D6-brane. In particular, 
\begin{equation}
    Q_6 = \frac{1}{2\pi} \bigl (N_{\mathrm{D}6} \mu_{\mathrm{D}6} + N_{\mathrm{O}6} \mu_{\mathrm{O}6} \bigr ) \quad.
\end{equation}
Working, for simplicity, in the instance such that
\begin{equation}
T_{\mathrm{D}6} = \mu_{\mathrm{D}6} \,=\, 2\pi \quad, \quad T_{\mathrm{O}6^-} = \mu_{\mathrm{O}6^-} \,=\, -4\pi \quad,
\end{equation}
and making the corresponding fluxes explicit, \eqref{eq:modifiedtadpole} reads
\begin{equation} 
  a_3 b_0 - 3 a_2 c_1 = (N_{\mathrm{D}6} -2 N_{\mathrm{O}6^-} \bigr )
\end{equation}
hence
\begin{equation} \label{eq:VDBIzero}
    V^{(0)}_{\mathrm{DBI}} = \sigma^3 \tau^{-3} \bigl (\tilde{{F}}_{(0)} \tilde{b}_0 - 3 \tilde{f}_2 c_1 \bigr) \quad.
\end{equation}
The zeroth order in $\lambda$ of the potential arising from the gauged supergravity is fully recovered by using in \eqref{eq:VDBIzero} the WZ-modified fluxes \eqref{eq:Mod_F0}, \eqref{eq:Mod_f2} and the modified metric flux
\begin{equation}
    \tilde{b}_0 = b_0 - 3 b_1 \chi_2 \quad.
\end{equation}
This arises from the definition $H_{(3)} = \dd B_{(2)}$ and corresponds to all terms in the potential scaling as $\tau^{-2} \rho^{-3} \sigma^{6}$ (also see \eqref{eq:VH3} and Table \ref{Table:Scaling_of_fluxes}).
Let us notice that $b_1$ and $c_1$ fluxes do not get corrections by the WZ and DBI actions, as the potential has no other terms with the same scaling as those in the  \eqref{eq:Vomega} and quadratic in these fluxes. This is why in \eqref{eq:VDBIzero} the flux $c_1$ is not modified.

At further order in $\lambda$ instead, only D6-branes can give rise to non-trivial terms in the DBI action. Due to \eqref{eq:sqrtM} and \eqref{eq:sqrtQ}, one finds that 
\begin{equation}
   V^{(2)}_{\mathrm{DBI}} = \frac{\sigma^3}{\tau^3} \mathrm{Tr} \Bigl [\frac{1}{2} \rho^{-1} \sigma^{-2} {{\mathbb{M}^{(2)}}^a}_{a} - \frac{1}{4} \rho^{-2} \sigma^{-4} {{\mathbb{M}^{(1)}}^a}_{b} {{\mathbb{M}^{(1)}}^b}_{a} - \frac{1}{4} {A^i}_j {A^j}_i \Bigr ]\quad,
\end{equation}
with the trace meant on the gauge generators as in \eqref{eq:GeneratorRelations}. 

If we consider all terms arising from reduction of the DBI action \eqref{eq:VDBIsecond} up to the second order in $\lambda$ and we add them to the bulk action $V_{\textup{Bulk}}$ \eqref{eq:VBulk} with all the field strengths replaced with the modified ones, we get the full potential of our theory, which perfectly matches with the scalar potential \eqref{eq:GaugedSugraPotential} coming from the $4D$ gauged supergravity description, once we consider the proper dictionary between scalar degrees of freedom in the $\mathrm{SO}(3)$ truncation \eqref{eq:dictionary_chi}, \eqref{eq:dictionary_phi}.

\section{Analysis of vacuum solutions} \label{section:Vacua}
Vacuum solutions are found by imposing the extremality condition of the potential \eqref{eq:GaugedSugraPotential} with respect to the scalar fields of the theory (collectively named here as $\Sigma$)
\begin{equation} \label{eq:CriticPotential}
    \partial_\Sigma V(\Sigma)|_{\Sigma_0}= 0 \quad,
\end{equation}
where $\Sigma_0$ represents the corresponding vacuum expectation values. We work in the \textit{flux picture} \cite{Dibitetto:2011gm}, namely we fix one point (or a subset) in the scalar manifold and then we look for the (real) flux values that minimize the potential. These critical points also need to satisfy the quadratic constraints \eqref{eq:quadraticconstraints} arising from the consistency of the gauged supergravity formulation. The set of polynomials associated to equations \eqref{eq:CriticPotential} and \eqref{eq:quadraticconstraints}, always quadratic in the fluxes, generate a polynomial ideal $I$, which can be studied with the aid of algebraic geometry techniques, as in \cite{Gray:2006gn, Guarino:2008ik, Dibitetto:2011gm}. In particular, we exploit the Gianni-Trager-Zacharias (GTZ) algorithm \cite{GIANNI1988149} to identify the decomposition of $I$ into primary ideals $J_i$ 
\begin{equation*}
    I = J_1 \cap \dots \cap J_n \quad,
\end{equation*}
whose existence is ensured by the Lasker-Noether theorem. 
This task has been accomplished by making use of the \textsc{\,Singular} software \cite{DGPS}. If we denote with $\mathcal{Z}(I)$ the set of common zeros of polynomials in the ideal $I$, which is the same as the set of zeros of the generating polynomials, we always have
$$
\mathcal{Z} \Big( \bigcap_{i=1}^n J_i \Big) = \bigcup_{i=1}^n \mathcal{Z} (J_i) \quad.
$$
Therefore we can find the real roots of the ideal $I$ by finding independently those of its primary ideals $J_i$, or equivalently of the associated primes $\sqrt{J_i}$.

\subsection{Solutions in the origin}
To begin with, we look for vacuum solutions restricting to the origin of the scalar manifold, \emph{i.e.}, setting to zero both the open and the closed string scalars. 

This is just a starting point, because the approach of going to the origin is not completely general in our case: we cannot perform arbitrary transformations from the duality group that simultaneously rotate the embedding tensor and the point in the scalar manifold, since a generic rotation would not leave invariant the set of flux components that we have chosen to turn on.

Looking at the expression of the scalar potential in the origin, we can see that, among the open string fluxes, it does not depend on the structure constants of the Yang-Mills gauge group $\tensor{g}{_{IJ}^K}$, then the only possibility to find new vacua that solve the quadratic constraint \eqref{QC_open_N3} is to have a non-vanishing field strength $\tensor{\mathcal{F}}{^I_{ab}}$. Critical points of this kind are actually found: the only possible solutions are 8 1-dimensional extrema, which can be grouped into two families of four 1-parameter solutions each, differing only in the choice of two signs $s_1$, $s_2 = \pm 1$, as shown in table \ref{Solutions_origin}. 
 
\begin{table}[H]
\centering
\renewcommand{\arraystretch}{2}
\resizebox{\textwidth}{!}{
\begin{tabular}{|c||c|c|c|c|c|c|c|c|c|c|}
       \hline
        Solution & $a_0$ & $a_1$ & $a_2$ & $a_3$ & $b_0$ & $b_1$ & $c_0$ & $c_1 = \bar{c}_1$ & $g_0$ & $g_1$\\\hline\hline
        $\mathbf{A}$ & $\lambda$ & $0$ & $0$ & $\displaystyle{s_1 \frac{\sqrt{5} + 3 \sqrt{13}}{14} \lambda}$ & $0$ & $\displaystyle{-\frac{11 + \sqrt{65} }{14}\lambda}$ & $0$ & $- \lambda$  & $\displaystyle{s_2 \frac{5 \sqrt{10} +  \sqrt{26}}{14}\lambda}$ & $0$\\\hline
        $\mathbf{B}$ & $\lambda$ & $0$ & $0$ & $\displaystyle{s_1 \frac{\sqrt{5} - 3 \sqrt{13}}{14} \lambda}$ & $0$ & $\displaystyle{\frac{-11 + \sqrt{65} }{14}\lambda}$ & $0$ & $- \lambda$  & $\displaystyle{s_2 \frac{5 \sqrt{10} -  \sqrt{26}}{14}\lambda}$ & $0$\\\hline
\end{tabular}
}
\caption{\emph{The two families of solutions in the origin depending on two signs $s_1$, $s_2$.}}
\label{Solutions_origin}
\end{table}

The values of the scalar potential for these solutions are
\begin{equation}
    V_0 (\mathbf{A}) = -\frac{3 \big( 61 + 3 \sqrt{65} \big) }{1568}\lambda^2 \quad , \qquad V_0 (\mathbf{B}) = \frac{3 \big( -61 + 3 \sqrt{65}\big) }{1568}\lambda^2 \quad .
\end{equation}

One may ask now what is the amount of supersymmetry preserved at any of the critical points. To this purpose, we can introduce the gravitini mass matrix $\tensor{A}{_1^{ij}}$ and the fermionic shift matrix $\tensor{A}{_2^{ij}}$, defined as
\begin{equation}
\begin{split}
    \tensor{A}{_1^{ij}}&  = \varepsilon^{\alpha \beta} \; \big( \mathcal{V}_{\alpha}\big)^{*} \tensor{\mathcal{V}}{_{[kl]}^M}\tensor{\mathcal{V}}{_N^{[ik]}}\tensor{\mathcal{V}}{_P^{[jl]}} \tensor{f}{_{\beta M}^{NP}} \;,\\
    \tensor{A}{_2^{ij}}&  = \varepsilon^{\alpha \beta} \;  \mathcal{V}_{\alpha} \tensor{\mathcal{V}}{_{[kl]}^M}\tensor{\mathcal{V}}{_N^{[ik]}}\tensor{\mathcal{V}}{_P^{[jl]}} \tensor{f}{_{\beta M}^{NP}} \; \ .
    \label{eq:fermion-shift-A2}
\end{split}    
\end{equation}
Here the complexified $\mathrm{SL}(2,\mathbb{R})$ vielbein $\mathcal{V}_{\alpha}$ is written as
\begin{equation}
    \mathcal{V}_{\alpha}= e^{\phi/2} \big( \chi + i e^{-\phi}, 1 \big)
\end{equation}
and it is such that $M_{\alpha \beta} = \mathcal{V}_{\alpha} \big( \mathcal{V}_{\beta}\big)^{*} $, while the complexified vielbein $\tensor{\mathcal{V}}{_M^{[ij]}}$ is built from the $\mathrm{SO}(6,6+\mathfrak{N})$ vielbein rotated to Cartesian coordinates $\tensor{\mathring{\mathcal{V}}}{_M^{\underline{M}}}$ by taking its timelike components $\tensor{\mathring{\mathcal{V}}}{_M^{\underline{m}}}$ and using the $\mathrm{SO}(6)$ 't Hooft symbols to turn the $\mathrm{SO}(6)$ index into a couple of antisymmetric $\mathrm{SU}(4)$ indices. 
$\tensor{A}{_1^{ij}}$ and $\tensor{A}{_2^{ij}}$ satisfy the following Killing spinor equations \cite{Schon:2006kz}
\begin{equation} \label{eq:Killing_spinor}
    \tensor{A}{_1^{ij}} q_j = \sqrt{-3 V} q^i \;, \qquad q_j \tensor{A}{_2^{ji}}=0 \;.
\end{equation}
In the case of our $\mathrm{SO}(3)$ truncation, the diagonalized gravitini mass matrix takes the form
\begin{equation}
    A_1 = \text{diag} ( \kappa_1, \kappa_2, \kappa_2 , \kappa_2 ) \;,
\end{equation}
due to the splitting of the $\mathrm{SU}(4)$ with respect to the action of $\mathrm{SO}(3)$, so that the possible amounts of supersymmetry preserved are $\mathcal{N}=4$, $\mathcal{N}=3$, $\mathcal{N}=1$ or $\mathcal{N}=0$ depending on whether the absolute values of $\kappa_1$ and $\kappa_2$ coincide with $\sqrt{-3 V}$.

Alternatively, the preserved supersymmetry can be investigated by finding the number of null eigenvalues of the fermionic shift matrix $A_2^{ij}$. In Table \ref{tab:SUSY-preserved} we show the necessary conditions on the fluxes for the solutions in the origin to preserve different amounts of supersymmetry. 

\begin{table}[t]
\centering
\begin{tabular}{|c|c|c|}
\hline
	$\N=1$ 	 & $\N=3$  & $\N=4$ \\ \hline \hline
	$\begin{aligned}
	a_0&= 3 a_2- 3 b_1+6 c_1-3 \bar{c}_1\\
	a_3&= 3 a_1-b_0- 3 c_0
	\end{aligned}$	& 
    $\begin{aligned}
    a_0&=-a_2+b_1-2 c_1-3\bar{c}_1\\
	a_3&=-a_1-b_0+c_0
    \end{aligned}$  &
    $\begin{aligned} 			a_0&=-3\bar{c}_1,\quad &a_1=c_0\\
	a_2&=b_1-2c_1,\quad &a_3=b_0
    \end{aligned}$ \\ \hline
\end{tabular}
\caption{\emph{Constraints on flux configurations for admitting solutions that preserve some supersymmetry at the origin.}}
\label{tab:SUSY-preserved}
\end{table}

\begin{table}[t]
\centering
\begin{tabular}{|c||c|c|c|c|c|c|c|c|c|}
\hline
    Solution &	$a_0$ &	$a_1$  &	$a_2$ 	 &  	$a_3$	&	$b_0$	& $b_1$	&$ c_0$ &	$c_1=\bar{c}_1$  	&	$g_{1}$	 \\ \hline \hline
    $\bold{1}_{(s_1)}$ &$\frac{3\sqrt{10}}{2}\lambda$ & $s_1\frac{\sqrt{6}}{2}$ &	 $-\frac{\sqrt{10}}{6}\lambda$	 & 	$s_1\frac{5\sqrt{6}}{6}\lambda$	&	$-s_1\frac{\sqrt{6}}{3}\lambda$	&		$\frac{\sqrt{10}}{3}\lambda$ & $s_1\frac{\sqrt{6}}{3}$&	$\sqrt{10}\lambda$ &	$g_{1}$	\\\hline
\end{tabular}
\caption{\emph{$\N=1$ vacua solutions at the origin ($s_1=\pm1$).}}
\label{tab:N1-solution}
\end{table}


According to the spinor equations \eqref{eq:Killing_spinor}, or equivalently to the conditions in table \ref{tab:SUSY-preserved}, all the solutions in families $\mathbf{A}$ and $\mathbf{B}$ are non-supersymmetric.
Indeed, requiring a non-trivial vacuum in the origin which partially preserves supersymmetry gives us back 2 solutions with residual $\mathcal{N}=1$ having vanishing $g_0$ and free value for $g_1$, which are the immediate generalization of the solutions already appearing in \cite{Dibitetto:2011gm}, \cite{Danielsson:2017max} (named as family $1$ with choice of the sign $s_2=+1$ and arbitrary $s_1$), that we report in Table \ref{tab:N1-solution} for completeness.

In order to understand the stability properties of each solution, it is not enough to determine the masses of the $\mathrm{SO}(3)$-invariant scalar fields from second-order derivatives of the scalar potential. Indeed, a critical point could be made unstable by fluctuations of any of the $\mathcal{N}=4$ scalars, including those ones that are set to zero by the truncation. The computation of the full mass spectra for any of the solutions can be performed adopting the formulation of \cite{Borghese:2010ei}. Stability of an $\text{AdS}_4$ solutions is established if all the mass eigenvalues satisfy the B. F. bound given by
\begin{equation}
    m^2 \geq - \frac{3}{4} \;.
\end{equation}
In the case of solutions of type $\mathbf{A}$ and $\mathbf{B}$, the masses do not depend on the signs $s_1$, $s_2$ and for both families the lowest eigenvalue does not satisfy the B.F. bound, then all these critical points are not stable. Further details about the complete set of normalized masses can be found in the appendix \ref{appendix_masses}.

\subsection{Out of the origin}
Moving out of the origin of the moduli space, in particular allowing
the open string scalars to have a non-vanishing value, we find 16 new critical points of the potential, all placed away from the origin only along the $\mathcal{A}$ direction. As in the previous case, they can be arranged into families of 4 solutions depending on two signs $s_1$, $s_2$. 

\begin{table}[H]
\centering
\renewcommand{\arraystretch}{2}
\resizebox{\textwidth}{!}{
\begin{tabular}{|c||c|c|c|c|c|c|c|c|c|c|}
        \hline
        Solution & $a_0$ & $a_1$ & $a_2$ & $a_3$ & $b_0$ & $b_1$ & $c_0$ & $c_1\,,\,\bar{c}_1$ & $g_0$ & $g_1$\\\hline\hline
        $\mathbf{1}$ & $\displaystyle{\Big(s_2 \frac{3}{2} - \frac{1}{2} \mathcal{A} ^2 \Big) \lambda}$ & $\displaystyle{s_1 \frac{1}{2} \sqrt{\frac{3}{5}} \lambda}$ & $\displaystyle{-s_2 \frac{\lambda}{6}}$ & $\displaystyle{s_1 \frac{1}{2} \sqrt{\frac{5}{3}} \lambda}$ & $\displaystyle{-s_1 s_2 \frac{\lambda}{\sqrt{15}}} $ & $\displaystyle{\frac{\lambda}{3}}$ & $\displaystyle{s_1 s_2 \frac{\lambda}{\sqrt{15}}}$ & $\displaystyle{\lambda}$ & $0$ & $\displaystyle{-\frac{\lambda}{\mathcal{A}}}$ \\ [2pt]\hline
        $\mathbf{2}$ & $\displaystyle{\Big(s_2 \frac{5}{3} - \frac{1}{2} \mathcal{A} ^2 \Big) \lambda}$ & $0$ & $0$ & $\displaystyle{s_1 \frac{\sqrt{5}}{3}\lambda}$ & $0$ & $\displaystyle{\frac{\lambda}{3}}$ & $0$ & $\displaystyle{\lambda}$ & $0$ & $\displaystyle{-\frac{\lambda}{\mathcal{A}}}$ \\[2pt]\hline
        $\mathbf{3}$ & $\displaystyle{\Big(s_2 - \frac{1}{2} \mathcal{A} ^2 \Big) \lambda}$ & $\displaystyle{- s_1 \frac{\lambda}{\sqrt{3}}}$ & $\displaystyle{s_2 \frac{\lambda}{3}}$ & $\displaystyle{ s_1 \frac{\lambda}{\sqrt{3}}}$ & $\displaystyle{s_1 s_2 \frac{\lambda}{\sqrt{3}}} $ & $\displaystyle{\frac{\lambda}{3}}$ & $\displaystyle{- s_1 s_2 \frac{\lambda}{\sqrt{3}}}$ & $\displaystyle{\lambda}$ & $0$ & $\displaystyle{-\frac{\lambda}{\mathcal{A}}}$ \\[2pt]\hline
        $\mathbf{4}$ & $\displaystyle{\Big(s_2 \sqrt{5} - \frac{1}{2} \mathcal{A} ^2 \Big) \lambda}$ & $0$ & $0$ & $\displaystyle{s_1 \lambda}$ & $0$ & $\lambda$ & $0$ & $\lambda$ & $0$ & $\displaystyle{-\frac{\lambda}{\mathcal{A}}}$ \\[2pt]\hline
\end{tabular}
}
\caption{\emph{The four families of solutions depending on two signs $s_1$, $s_2$.}}
\label{Solutions_4families}
\end{table}

As shown in table \ref{Solutions_4families}, the values of the fluxes for these solutions can be expressed in terms of two parameters, one of them being precisely the value of the scalar field $\mathcal{A}$. Indeed, they can all be interpreted as a generalization of the 1-dimensional solutions discussed in \cite{Dibitetto:2011gm} and \cite{Danielsson:2017max} in a closed string scenario. Those solutions can be restored if we take the limit $\mathcal{A} \, \to \, 0$, with the condition that $\frac{\lambda}{\mathcal{A}}$ remains constant: in this way, the $\mathcal{A}^2$ correction to the flux $a_0$ disappears and the flux $\tensor{g}{_{IJ}^K}$ becomes independent of the closed string sector. In other words, we come back to a solution in the origin, with an open string flux $\tensor{g}{_{IJ}^K}$ that, even if not vanishing, does never contribute to the scalar potential, as already explained.  

The vacuum energy densities associated to these solutions are
\begin{equation}
    V_0 (\mathbf{1}) = -\frac{\lambda^2}{10} \quad , \quad V_0 (\mathbf{2}) = -\frac{5 \lambda^2}{48}  \quad , \quad V_0 (\mathbf{3}) = -\frac{\lambda^2}{12}  \quad , \quad V_0 (\mathbf{4}) = -\frac{3 \lambda^2}{16}  \quad ,
\end{equation}
which coincide, up to a proper rescaling of the parameter $\lambda$ employed, with the values for the old solutions in the closed string sector.

Looking for further new solutions possibly out of the origin, we focused in particular on the possibility of vacua with no Roman's mass, namely $a_3 = 0$. However, this investigation led us to the conclusion that - besides vacua with all open string fluxes vanishing - the only such vacua are Minkowski vacua in the origin with fluxes
\begin{equation}
    b_1=a_2 \;, \qquad  a_0=a_1=a_3=b_0=c_0=c_1=\bar{c}_1=0 \;, \qquad g_0=0 \;,
\end{equation}
for any value of the flux $g_1$. Taking into account the fact that $g_1$ is not coupled to the other fluxes and does not contribute to the scalar potential, these vacua correspond to a 1-dimensional subset of the 2-dimensional critical points described in \cite{Dibitetto:2011gm} in a purely closed string setting, dual to GKP ones \cite{Giddings:2001yu}.

As far as residual supersymmetry is concerned, solving the eigenvalues equation of $\tensor{A}{_1^{ij}}$, we find that only the solutions of family $\mathbf{1}$ with $s_2=+$ preserve $\mathcal{N}=1$ supersymmetry, having $\kappa_1=\sqrt{-3 V}$, while the remaining ones are non-supersymmetric. We can have a double check of this result by diagonalizing the fermionic shift matrix $\tensor{A}{_2^{ij}}$: only one zero eigenvalue appears in the case of solutions of family $\mathbf{1}$ with $s_2=+$, while all the other solutions give a matrix $\tensor{A}{_2^{ij}}$ with non-vanishing eigenvalues.

In order to understand the stability properties of each solution, we would like to compute the associated mass spectra for the complete set of scalar fields. To this purpose, we can still follow the procedure developed in \cite{Borghese:2010ei}, but only after properly rotating the embedding tensor. Since the procedure is only valid at the origin, we have to find the rotation that brings a generic point of the moduli space with non-vanishing $\mathcal{A}$ back to the origin and apply the inverse rotation to the embedding tensor indices. 

Tables \ref{Masses_s2Plus} and \ref{Masses_s2Minus} collect the results obtained for the normalized masses. In each table, we colour the multiplicities associated to the masses coming from the open string sector; in particular, the masses that are common to any choice of the signs are shown in blue, while the values that depend on the sign $s_2$ are highlighted in red.

We can deduce that the stable extrema of the scalar potential, i.e. the solutions where tachyons are not present or satisfy the B.F. bound $m^2 \geq - \frac{3}{4}$, are
\begin{itemize}
    \item solutions of family $\mathbf{1}$ for any choice of the signs;
    \item solutions of family $\mathbf{3}$ for any choice of the signs;
    \item solutions of family $\mathbf{4}$ only for $s_2=+$.
\end{itemize}
In the remaining cases, there is a tachyon exceeding the B.F. bound, which is highlighted in gray in tables \ref{Masses_s2Plus} and \ref{Masses_s2Minus}.

\begin{table}
\begin{minipage}{0.6\linewidth}
\centering
\(\begin{array}{cc}
    \toprule
        \multicolumn{2}{c}{\mathbf{1}\, , \, s_2=+1}\\[3pt]
    \midrule
        -\frac{2}{3} & 1+ \textcolor{red}{1} \\[4pt]
        0 & 10+ \textcolor{blue}{4} \\[4pt]
        \frac{1}{3} \big( 4-\sqrt{6} \big) & 1 \\[4pt]
        \frac{1}{18} \Big( 89 -5 \sqrt{145} - \sqrt{606- 30 \sqrt{145}}\Big) & 5 \\[4pt]
        \frac{11}{6} & \textcolor{red}{3} \\[4pt]
        \frac{1}{3} \big(4+\sqrt{6}\big) & 1 \\[4pt]
        \frac{1}{18} \Big( 89 -5 \sqrt{145} + \sqrt{606- 30 \sqrt{145}}\Big) & 5 \\[4pt]
        \frac{29}{9} & 3 \\[4pt]
        \frac{1}{9} \big(47-\sqrt{159}\big) & 1 \\[4pt]
        \frac{1}{6} \big(43-\sqrt{94}\big) & \textcolor{red}{5} \\[4pt]
        \frac{1}{18} \Big( 89 +5 \sqrt{145} - \sqrt{606+ 30 \sqrt{145}}\Big) & 5 \\[4pt]
        \frac{1}{9} \big(47+\sqrt{159}\big) & 1 \\[4pt]
        \frac{1}{6} \big(43+\sqrt{94}\big) & \textcolor{red}{5} \\[4pt]
        \frac{1}{18} \Big( 89 +5 \sqrt{145} + \sqrt{606+ 30 \sqrt{145}}\Big) & 5 \\[4pt]
    \bottomrule
\end{array}\)   
\end{minipage}
\begin{minipage}{0.35\linewidth}
\centering
\(\begin{array}{cc}
    \toprule
        \multicolumn{2}{c}{\mathbf{2}\, , \, s_2=+1}\\[3pt]
    \midrule
        \colorbox{lightgray}{$-\frac{4}{5}$} & 1 \\[4pt]
        -\frac{2}{5} & 1 \\[4pt]
        0 & 10+ \textcolor{blue}{4} \\[4pt]
        \frac{4}{5} & \textcolor{red}{5} \\[4pt]
        \frac{1}{15} \big( 77- 5\sqrt{145} \big) & 5 \\[4pt]
        2 & 1+ \textcolor{blue}{3} \\[4pt]
        \frac{2}{15} \big( 31- \sqrt{145} \big) & 5 \\[4pt]
        \frac{46}{15} & 3 \\[4pt]
        \frac{16}{5} & \textcolor{red}{1} \\[4pt]
        \frac{64}{15} & 1 \\[4pt]
        \frac{2}{15} \big( 31+ \sqrt{145} \big) & 5 \\[4pt]
        6 & \textcolor{blue}{5} \\[4pt]
        \frac{20}{3} & 1 \\[4pt]
        \frac{1}{15} \big( 77+ 5\sqrt{145} \big) & 5 \\[4pt]
    \bottomrule
\end{array}\)   
\end{minipage}

\vspace{30pt}

\begin{minipage}{0.60\linewidth}
\centering
\(\begin{array}{cc}
    \toprule
        \multicolumn{2}{c}{\mathbf{3}\, , \, s_2=+1}\\[3pt]
    \midrule
        0 & 11+ \textcolor{blue}{4} + \textcolor{red}{1} \\[4pt]
        2 & 2 \\[4pt]
        \frac{1}{3} \big( 19-\sqrt{145} \big) & 10 \\[4pt]
        3 & \textcolor{blue}{3} \\[4pt]
        \frac{14}{3} & 3 \\[4pt]
        \frac{20}{3} & 2 \\[4pt]
        9 & \textcolor{blue}{5} + \textcolor{red}{5} \\[4pt]
        \frac{1}{3} \big( 19+\sqrt{145} \big) & 10 \\[4pt]
    \bottomrule
\end{array}\)   
\end{minipage}
\begin{minipage}{0.35\linewidth}
\centering
\(\begin{array}{cc}
    \toprule
        \multicolumn{2}{c}{\mathbf{4}\, , \, s_2=+1}\\[3pt]
    \midrule
        \frac{4}{3} \big( 2 -\sqrt{5} \big) & \textcolor{red}{5} \\[4pt]
        0 & 16+ \textcolor{blue}{4} \\[4pt]
        \frac{2}{3} & \textcolor{blue}{3} \\[4pt]
        \frac{4}{3} & 6 \\[4pt]
        2 & 4+ \textcolor{blue}{5} \\[4pt]
        \frac{2}{3} \big( 1+\sqrt{5} \big) & \textcolor{red}{1} \\[4pt]
        \frac{8}{3} & 5 \\[4pt]
        6 & 6 \\[4pt]
        \frac{20}{3} & 1 \\[4pt]
    \bottomrule
\end{array}\)   
\end{minipage}

\caption{\emph{Normalized masses for the four families of solutions with the sign $s_2=+1$. Tachyons are highlighted with a gray background. The multiplicities associated to the masses coming from the open string sector are coloured (blue for masses that are common to any choice of the signs, red for those depending on $s_2$).}}
\label{Masses_s2Plus}
\end{table}

\begin{table}
\begin{minipage}{0.60\linewidth}
\centering
\(\begin{array}{cc}
    \toprule
        \multicolumn{2}{c}{\mathbf{1}\, , \, s_2=-1}\\[3pt]
    \midrule
        -\frac{2}{3} & 1 \\[4pt]
        0 & 10+ \textcolor{blue}{4} \\[4pt]
        \frac{1}{3} \big( 4-\sqrt{6} \big) & 1 \\[4pt]
        \frac{1}{18} \Big( 89 -5 \sqrt{145} - \sqrt{606- 30 \sqrt{145}}\Big) & 5 \\[4pt]
        \frac{5}{6} & \textcolor{red}{5} \\[4pt]
        \frac{1}{3} \big(4+\sqrt{6}\big) & 1 \\[4pt]
        \frac{1}{18} \Big( 89 -5 \sqrt{145} + \sqrt{606- 30 \sqrt{145}}\Big) & 5 \\[4pt]
        \frac{5}{2} & \textcolor{red}{3} \\[4pt]
        \frac{29}{9} & 3 \\[4pt]
        \frac{10}{3} & \textcolor{red}{1} \\[4pt]
        \frac{1}{9} \big(47-\sqrt{159}\big) & 1 \\[4pt]
        \frac{1}{18} \Big( 89 +5 \sqrt{145} - \sqrt{606+ 30 \sqrt{145}}\Big) & 5 \\[4pt]
        \frac{1}{9} \big(47+\sqrt{159}\big) & 1 \\[4pt]
        \frac{15}{2} & \textcolor{red}{5} \\[4pt]
        \frac{1}{18} \Big( 89 +5 \sqrt{145} + \sqrt{606+ 30 \sqrt{145}}\Big) & 5 \\[4pt]
    \bottomrule
\end{array}\)   
\end{minipage}
\begin{minipage}{0.35\linewidth}
\centering
\(\begin{array}{cc}
    \toprule
        \multicolumn{2}{c}{\mathbf{2}\, , \, s_2=-1}\\[3pt]
    \midrule
        \colorbox{lightgray}{$-\frac{4}{5}$} & 1+\textcolor{red}{1} \\[4pt]
        -\frac{2}{5} & 1 \\[4pt]
        0 & 10+ \textcolor{blue}{4} \\[4pt]
        \frac{1}{15} \big( 77- 5\sqrt{145} \big) & 5 \\[4pt]
        2 & 1+ \textcolor{blue}{3} \\[4pt]
        \frac{2}{15} \big( 31- \sqrt{145} \big) & 5 \\[4pt]
        \frac{46}{15} & 3 \\[4pt]
        \frac{64}{15} & 1 \\[4pt]
        \frac{2}{15} \big( 31+ \sqrt{145} \big) & 5 \\[4pt]
        6 & \textcolor{blue}{5} \\[4pt]
        \frac{20}{3} & 1 \\[4pt]
        \frac{44}{5} & \textcolor{red}{5} \\[4pt]
        \frac{1}{15} \big( 77+ 5\sqrt{145} \big) & 5 \\[4pt]
    \bottomrule
\end{array}\)   
\end{minipage}

\vspace{30pt}

\begin{minipage}{0.60\linewidth}
\centering
\(\begin{array}{cc}
    \toprule
        \multicolumn{2}{c}{\mathbf{3}\, , \, s_2=-1}\\[3pt]
    \midrule
        0 & 11+ \textcolor{blue}{4} \\[4pt]
        1 & \textcolor{red}{5} \\[4pt]
        2 & 2 \\[4pt]
        \frac{1}{3} \big( 19-\sqrt{145} \big) & 10 \\[4pt]
        3 & \textcolor{blue}{3} \\[4pt]
        4 & \textcolor{red}{1} \\[4pt]
        \frac{14}{3} & 3 \\[4pt]
        \frac{20}{3} & 2 \\[4pt]
        9 & \textcolor{blue}{5} \\[4pt]
        \frac{1}{3} \big( 19+\sqrt{145} \big) & 10 \\[4pt]
    \bottomrule
\end{array}\)   
\end{minipage}
\begin{minipage}{0,35\linewidth}
\centering
\(\begin{array}{cc}
    \toprule
        \multicolumn{2}{c}{\mathbf{4}\, , \, s_2=-1}\\[3pt]
    \midrule
        \colorbox{lightgray}{$\frac{2}{3} \big( 1 -\sqrt{5}\big)$}  & \textcolor{red}{1} \\[4pt]
        0 & 16+ \textcolor{blue}{4} \\[4pt]
        \frac{2}{3} & \textcolor{blue}{3} \\[4pt]
        \frac{4}{3} & 6 \\[4pt]
        2 & 4+ \textcolor{blue}{5} \\[4pt]
        \frac{8}{3} & 5 \\[4pt]
        \frac{4}{3} \big( 2+\sqrt{5} \big) & \textcolor{red}{5} \\[4pt]
        6 & 6 \\[4pt]
        \frac{20}{3} & 1 \\[4pt]
    \bottomrule
\end{array}\)   
\end{minipage}

\caption{\emph{Normalized masses for the four families of solutions with the sign $s_2=-1$. Tachyons are highlighted with a gray background. The multiplicities associated to the masses coming from the open string sector are coloured (blue for masses that are common to any choice of the signs, red for those depending on $s_2$).}}
\label{Masses_s2Minus}
\end{table}

\subsection{The case $\mathfrak{N}=6$ at the origin}

We have also explored the case $\mathfrak{N}=6$ at the origin and have found some solutions beyond the ones that can be trivially extended from the configurations found for $\mathfrak{N}=3$. 

Having $\mathfrak{N}=6$ allows us to consider, at least, 2 nonvanishing values for the structure constants (modulo their antisymmetric components). This way, we can solve the quadratic constraint \eqref{eq:QC-nonsemisimple} by considering gauge fluxes ${\cal F}^I{}_{ab}$ along the orthogonal components for which the structure constant are nonzero. 

By splitting the indices $I$ as $I=\{I',\, I''\}$ ($I',I''=1,2,3$) and for consistency with the SO(3) truncation, we consider the following fluxes:
\begin{align}
    {\cal F}^{I'}{}_{ab} =&\ \varepsilon_{abc} \delta^{c I'} g_0 \ ,
    &
    g_{I'J'}{}^{K'}=&\ \varepsilon_{I'J'L'}\delta^{L'K'} g_1 \ ,
    \\
    {\cal F}^{I''}{}_{ab} =&\ \varepsilon_{abc} \delta^{c I''} g_0' \ ,
    &
    g_{I''J''}{}^{K''}=&\ \varepsilon_{I''J''L''}\delta^{L''K''} g_1' \ ,
\end{align}
which correspond  to $G_{\text{YM}}=\text{SO}(3)\times \text{SO}(3)$. 

Upon focusing on non-Minkowskian vacua and imposing the coexistence of both fluxes ${\cal F}^I{}_{ab}$ and $g_{IJ}{}^K$ via the constraints
\begin{align}
    g_0 \, g_1'\neq0 \ 
    \qquad\text{or}\qquad
    g_1 \, g_0'\neq0 \ ,
\end{align}
we obtain a set of 1-parameter solutions, which are shown in Table \ref{tab:N6-solutions}.

\begin{table}[t]
\centering
\renewcommand{\arraystretch}{1.6}
\begin{tabular}{|c||c|c|c|c|c|c|c|c|}
\hline
    Solutions &	$a_0$ 	 &  	$a_3$	&	$b_1$	&		$c_1=\bar{c}_1$  	&	$g_{1}$	& $ g'_{1}$	& $g_{0}$	& $g'_{0}$  \\ \hline \hline
    $\bold{1}_{(s_1,s_2)}$ &	 $\frac{\sqrt{65}-11}{4}\lambda$	 & 	$s_1\frac{\sqrt{13}-\sqrt{5}}{2}\lambda$	&	$\lambda$	&		$-\frac{\sqrt{65}-11}{4}\lambda$ 	&	$g_{1}$	& $ 0$	& $0$	&  $s_2\frac{3\sqrt{5}-\sqrt{13}}{2\sqrt{2}}\lambda$\\\hline
    $\bold{2}_{(s_1,s_2)}$ &	 $\frac{\sqrt{65}+11}{4}\lambda$	 & 	$s_1\frac{\sqrt{13}+\sqrt{5}}{2}\lambda$	&	$\lambda$	&		$-\frac{\sqrt{65}+11}{4}\lambda$ 	&	$g_{1}$	& $ 0$	& $0$	&  $s_2\frac{3\sqrt{5}+\sqrt{13}}{2\sqrt{2}}\lambda$\\\hline
    $\bold{1'}_{(s_1,s_2)}$ &	 $\frac{\sqrt{65}-11}{4}\lambda$	&	$s_1\frac{\sqrt{13}-\sqrt{5}}{2}\lambda$	&	$\lambda$	&	$-\frac{\sqrt{65}-11}{4}\lambda$ 	&	$0$	& $ g'_{1}$	&  $s_2\frac{3\sqrt{5}-\sqrt{13}}{2\sqrt{2}}\lambda$	& $0$	\\\hline
    $\bold{2'}_{(s_1,s_2)}$ &	 $\frac{\sqrt{65}+11}{4}\lambda$	&	$s_1\frac{\sqrt{13}+\sqrt{5}}{2}\lambda$	&	$\lambda$	&	$-\frac{\sqrt{65}+11}{4}\lambda$ 	&	$0$	& $ g'_{1}$	&  $s_2\frac{3\sqrt{5}+\sqrt{13}}{2\sqrt{2}}\lambda$	& $0$	\\\hline
\end{tabular}
\caption{\emph{Case $\mathfrak{N}=6$.  The 1-parameter families of vacuum solutions at the origin with  $s_i=\pm 1$. The coupling constants $g_1$ \& $g_1'$ remain arbitrary and decoupled from the dynamics. }}
\label{tab:N6-solutions}
\end{table}

The value of the potential at the origin for each family of solutions is:
\begin{align}
V(\bold{1}_{(s_1,s_2)})&=V(\bold{1'}_{(s_1,s_2)})=-\frac{3}{32}\left(9-\sqrt{65}\right)\lambda^2 \ < \ 0 \ ,\\
V(\bold{2}_{(s_1,s_2)})&=V(\bold{2'}_{(s_1,s_2)})=-\frac{3}{32}\left(9+\sqrt{65}\right)\lambda^2 \ < \ 0 \ ,
\end{align}
thus yielding AdS solutions in all cases. However, the minimum of the mass spectrum for $\bold{1}_{(s_1,s_2)}$ and $\bold{1'}_{(s_1,s_2)}$ is $m^2=-0,9997$, and for $\bold{2}_{(s_1,s_2)}$ and $\bold{2'}_{(s_1,s_2)}$ is $m^2=-1.3150$. So, none of them satisfy the B.F. bound, and therefore these solutions are unstable.

\section{Flux quantization and perturbative control}
\label{section:PerturbativeControl}
It is well-known that 10D supergravity is a good low-energy effective description for strings in the limit where both the higher-loop expansion governed by the string coupling $g_s$ and the higher-derivative expansion governed by $\alpha'$ are under control. In practice this means to adopt a compactification setup where the internal volume is large w.r.t. the string scale and $e^\Phi$ is small. 

Now, within the 4D supergravity description, if we restrict to the metric and the scalars, the action may be cast in the following form
\begin{equation}
    S_{\textrm{4D}} \,=\, \int d^4x \sqrt{-g_4}\,\left(\frac{M_{\textrm{Pl}}^2}{2}\left(\mathcal{R}_4-\frac{1}{2}K_{AB}\,\partial\phi^A\partial\phi^B\right)-M_{\textrm{Pl}}^4\, V(\phi)\right)\,+\,\dots \ ,
\end{equation}
where $K_{AB}$ is the metric on the scalar manifold and $M_{\textrm{Pl}}$ is the 4D Planck scale. Note that, when expressing things in this way all the fields and embedding tensor deformations appearing in the 4D action are \emph{dimensionless}. This means that, in order to make our 10D/4D dictionary dimensionally correct, we need to reinsert the right powers of $(2\pi\ell_s)$ everywhere in our reduction \emph{Ansatz}. This will read
\begin{equation}
    ds_{10}^2 \,=\, \tau^{-2} ds_4^2 \,+\, \left(2\pi\ell_s\right)^2\rho \left(\sigma^2 ds_{3,a}^2 \,+\, \sigma^{-2} ds_{3,i}^2\right) \ ,
\end{equation}
where $\rho$, $\tau$ and $\sigma$ are dimensionless. In particular, $\rho$ measures the internal volume in string units. In these units, the integrated fluxes need to be quantized according to
\begin{equation}
     \oint\limits_{\mathcal{C}_3}H_{(3)} \,\in\, \left(2\pi\ell_s\right)^2 \mathbb{Z}   \ , \qquad \oint\limits_{\mathcal{C}_p}F_{(p)} \,\in\, \left(2\pi\ell_s\right)^{p-1} \mathbb{Z} \ , \qquad \omega_{mn}{}^p \,\in\,  \mathbb{Z} \ .
\end{equation}
In this way, the objects appearing in our dictionary in table \ref{Table:Dictionary_fluxes_embedding_closedsector} are directly to be identified with the integers appearing above. For what concerns the open string flux, this obeys
\begin{equation}
     \oint\limits_{\mathcal{C}_2}\mathcal{F} \,\in\, 2\pi \mathbb{Z}   \ .
\end{equation}

The vacua presented here are all at $\rho=1$ and with unquantized fluxes. However, we will show in this section that, we can act with a 4D rescaling that transforms both the scalars and the fluxes into a configuration with physically desirable features. The rescaling we need is a combination 
\begin{equation}
    \mathbb{R}^+_{\Omega} \,\equiv\,\mathbb{R}^+_{\textrm{trombone}} \,\times\, \mathbb{R}^+_{\rho} \,\times\, \mathbb{R}^+_{\tau} \ ,
\end{equation}
where the trombone $\mathbb{R}^+$ rescales the whole 4D Lagrangian leaving the EOM's invariant.
\begin{equation}
    \begin{array}{lcclcclc}
     \rho \ \sim \ \Omega^{2}    &  ,  & & \tau \ \sim \ \Omega^{6} & , & & \sigma \ \sim \ \Omega^{0}  & ,
    \end{array}
\end{equation}
where $\Omega \, \in \, \mathbb{R}^+$ will be chosen to be parametrically large. Note that we do not want $\sigma$ to scale non-trivially in $\Omega$, as it appears in the 10D metric both with positive and negative powers, hence we would necessarily have some internal cycle dropping below the string scale when $\Omega$ is chosen large.
In table \ref{Table:Omega_fluxes} we display the scaling behavior of all the fluxes considered in our solutions.
\begin{table}[h!]
\renewcommand{\arraystretch}{1}
\begin{center}
\scalebox{1}[1]{
\begin{tabular}{|c||c|c|c|c|}
\hline
Fluxes & $H_{(3)}$ & $F_{(p)}$ & $\omega$ & $\mathcal{F}^I$\\[2mm]
\hline \hline
$\Omega$ weights & $\Omega^2$ & $\Omega^{2+p}$ & $\Omega^0$ & $\Omega^4$\\[2mm]
\hline
\end{tabular}
}
\end{center}
\caption{\it The scaling weights of all the fluxes that appear as ingredients in our massive type IIA compactification setup. } \label{Table:Omega_fluxes}
\end{table}
It may be worth mentioning that all the flux numbers that grow large in the large $\Omega$ regime, become insensitive to the flux quantization conditions.
Note that $\mathcal{F}^I$ contains both the pure YM flux part and its non-Abelian field strength contribution expressed in terms of the structure constants $g_{IJ}{}^{K}$, which need to scale as $\Omega^{-4}$ if $\mathcal{A}^I$ is assumed to scale like $\mathcal{F}^I$. This in turn implies that the gauge generators $t_I$ scale as $\Omega^{-4}$. In this way the full non-Abelian field strength $\mathcal{F}=\mathcal{F}^It_I$ is invariant under $\Omega$.

The physical scales controlling the reliability of our setup are analyzed in table \ref{Table:Omega_scales}, in terms of their parametric behavior in $\Omega$. As it may be seen in the table, all our AdS vacuum solutions presented in this work appear to be reliable, whether or not they preserve supersymmetry. However, they cannot be regarded as proper 4D vacua, in the sense that they fail to realize scale separation. As noted in \cite{Gautason:2015tig}, this is generically expected in 10D solutions with non-zero internal curvature. 
\begin{table}[h!]
\renewcommand{\arraystretch}{1}
\begin{center}
\scalebox{1}[1]{
\begin{tabular}{|c||c|c|c|c|}
\hline
Scales & $g_s$ & $\frac{Vol_6}{\left(2\pi\ell_s\right)^6}$ & $\frac{|\Lambda|}{M_{\textrm{Pl}}^4}$ & $\frac{\ell_{\textrm{KK}}}{\ell_{\textrm{AdS}}}$\\[2mm]
\hline \hline
$\Omega$ weights & $\Omega^{-3}$ & $\Omega^{6}$ & $\Omega^{-14}$ & $\Omega^0$\\[2mm]
\hline
\end{tabular}
}
\end{center}
\caption{\it The scaling weights of the physical scales in these compactifications. It is apparent that the string coupling is parametrically small and the internal volume is large in a regime where $\Omega\gg1$. However, this cannot be combined with scale separation, as the KK scale is of the same order as the AdS scale. } \label{Table:Omega_scales}
\end{table}

What about higher-derivative corrections? Since the coupling of open strings to the closed string sector in the DBI action appears as a finite $\alpha'$ effect, one might be worried about higher-order corrections. The NLO corrections in the DBI contributions to the scalar potential come at $\mathcal{O}(\lambda^4)$, and are of degree four in the fluxes. A typical term of such a form is
\begin{equation}
    g_{1}^4\mathrm{Tr}\left(t_It_I\right)\mathcal{A}^8\rho^{-4}\tau^{-3}\sigma^{-5} \ \sim \ \Omega^{-18} \ ,
\end{equation}
which is indeed subleading w.r.t. the $\Omega^{-14}$ scaling of the LO potential terms as read off from table \ref{Table:Omega_scales}. We checked that the complete set of corrections to the DBI potential at this order only contains terms that scale as $\Omega^{-18}$.

When it comes to higher-derivative corrections to the bulk action, we do not expect anything before $\mathcal{O}(\alpha'^3)$, and these contain, \emph{e.g.} the $\mathcal{R}_{10}^{(4)}$ term. A correction to the potential arising from this higher-curvature contribution would scale like
\begin{equation}
\sqrt{g_6}\,\rho^3\tau^{-4} \left(\rho^{-1}\mathcal{R}_{6}\right)^4 e^{-2\Phi} \ \sim \ \Omega^{-20} \ ,
\end{equation}
which is even more suppressed in the large $\Omega$ regime. Finally, string loop corrections will be suppressed by extra powers of $g_s \,\sim\,\Omega^{-3}$.

Two general comments are due before concluding. The former one concerns the stability issue for the non-supersymmetric AdS vacua presented here. Though they appear perturbatively stable, one might expect there to be a non-perturbative mechanism that destabilizes these solutions, in line with the AdS swampland conjecture stated in \cite{Ooguri:2016pdq}. However, the concrete construction of a viable decay channel still remains to be investigated.

The latter issue is to do with our way of understanding the AdS/CFT correspondence, even in a supersymmetric setup. Generically, we think of it as a holographic duality between the closed string setup that engineers a supergravity AdS vacuum and the corresponding strongly coupled CFT describing the open strings living on the worldvolume of the branes. In our case, the very AdS vacuum relies on a (semiclassical) coupling between closed strings and the open strings living on the worldvolume of the spacetime filling sources supporting the solution. This extra complication would require a rethinking of the holographic correspondence in presence of such brane sources. At present, we cannot yet argue that these vacua admit a CFT dual.

\section*{Acknowledgements}

We would like to thank Ivano Basile, Carlo Maccaferri, Miguel Montero and Marco Scalisi for stimulating discussions.
The work of JRB is supported by Fundaci\'on S\'eneca, Agencia de Ciencia y Tecnolog\'ia de la Regi\'on de Murcia under grant 21472/FPI/20.
The work of JRB and JJFM is supported by the Spanish Ministerio de Innovaci\'on y Ciencia and Universidad de Murcia under grants PID2021-125700NA-C22 and E024-18, respectively.
The work of GD is partially supported by the INFN project ST\&FI.

\appendix

\section{A democratic formulation for massive IIA supergravity}
Type IIA supergravity with Roman's mass \cite{Romans:1985tz} admits a democratic formulation \cite{Bergshoeff:2001pv}. The bosonic sector thereof contains the usual NS-NS fields $\{G, B_{(2)}, \Phi\}$, whereas the R-R degrees of freedom are doubled $\{C_{2p + 1}\}_{p=0, 1, 2, 3, 4}$. The following pseudoaction can be used as a mnemonic to obtain the equations of motion in the boson part:
\begin{equation} \label{eq:pseudoaction}
S = \frac{1}{2 \kappa_{10}^2} \int \mathrm{d}^{10} x \ \sqrt{-G} \Bigl (e^{-2 \Phi} \bigl (\mathcal{R} + 4 (\partial \Phi)^2 - \frac{1}{12} |H_{(3)}|^2   \bigr ) - \frac{1}{4} \sum_{p=0}^5 \frac{|F_{(2 p)}|^2}{(2 p)!} \Bigr) 
\end{equation} 
The field strengths appearing in the pseudoaction \eqref{eq:pseudoaction} are given by
\begin{gather}
    H_{(3)} = \dd B_{2} \,, \quad
    H_{(7)} = \dd B_{6} \ + \ \sum_{p=0}^3 (-1)^k C_{(7 - 2 p)} \ \wedge \ \bigr ( F_{(2p)} + F_0 \wedge \frac{B_{(2)}^p}{p!}  \bigl) 
    \,,\\
    \quad F= \dd C + \dd B \wedge C + F_{(0)} \, e^{-B} \;.\label{eq:bosonic_field_strengths}
\end{gather}

Making the formal sums in \eqref{eq:bosonic_field_strengths} explicit, we can write the field strengths of the R-R fields as
\begin{align}
    & F_0 = m \quad, \label{eq:Romansmass}\\
    & F_{(2)} = \dd C_{(1)} - F_{(0)} B_{(2)} \quad, \label{eq:F2}\\
    & F_{(4)} = \dd C_{(3)} + H_{(3)} \wedge C_{(1)} + F_{(0)} \frac{B_{(2)}^2}{2!}\quad,  \label{eq:F4}\\
    & F_{(6)} = \dd C_{(5)} + H_{(3)} \wedge C_{(3)} - F_{(0)} \frac{B_{(2)}^3}{3!}  \quad, \label{eq:F6}\\
    & F_{(8)} = \dd C_{(7)} + H_{(3)} \wedge C_{(5)} + F_{(0)} \frac{B_{(2)}^4}{4!}  \quad,\\
    & F_{(10)} = \dd C_{(9)} + H_{(3)} \wedge C_{(7)} -F_{(0)} \frac{B_{(2)}^5}{5!}  \quad.
\end{align}
In particular, Roman's mass contribution is encoded in \eqref{eq:Romansmass},
with constant $m$.

Because of their definitions, the field strengths naturally satisfy the modified Bianchi identities
\begin{align}
     \dd H_{(3)} = 0 \quad, & \qquad  \dd H_{(7)} -\frac{1}{2} \sum_{p=0}^4 F_{(2p)} \wedge \star F_{(2p+2)} = 0 \quad, \\
    \dd F_{(0)} = 0 \quad, & \qquad  \dd F_{(2)} - H_{(3)} \wedge F_{(0)}  = 0 \quad,  \label{eq:modifiedBianchiF2} \\
    \dd F_{(4)} - H_{(3)} \wedge F_{(2)} = 0 \quad, & \qquad  \dd F_{(6)} - H_{(3)} \wedge F_{(4)} = 0 \quad, \\
    \dd F_{(8)} - H_{(3)} \wedge F_{(6)} = 0 \quad, & \qquad  \dd F_{(10)} = 0 \quad. 
\end{align}
The pseudoaction needs to be supplemented by the following duality relations 
\begin{equation} \label{eq:duality}
     F_{(0)} = \star F_{(10)} \;, \qquad  F_{(2)} = - \star F_{(8)} \;, \qquad  F_{(4)} = \star F_{(6)} \;, \qquad H_{(7)} = e^{-2 \Phi} \star H_{(3)} \;. 
\end{equation}
This way, the original number of degrees of freedom in the R-R sector is restored. Moreover, \eqref{eq:duality} maps the modified Bianchi identities to the equations of motion for the form fields in the NS-NS and the R-R sector:
\begin{align}
     \dd \bigl (e^{2 \Phi}  \star H_{(7)} \bigr ) = 0 \quad, & \qquad \dd \bigl ( e^{- 2 \Phi}  \star H_{(3)} \bigr ) - \frac{1}{2} \sum_{p=0}^4 F_{(2p)} \wedge \star F_{(2p+2)} = 0 \quad, \\
    \dd \star F_{(10)} = 0 \quad, & \qquad \dd \star F_{(8)} + H_{(3)} \wedge \star F_{(10)}  = 0 \quad,  \\
    \dd \star F_{(6)} + H_{(3)} \wedge \star F_{(8)} = 0 \quad, & \qquad \dd \star F_{(4)} + H_{(3)} \wedge \star F_{(6)} = 0 \quad, \\
    \dd \star F_{(2)} + H_{(3)} \wedge \star F_{(4)} = 0 \quad, & \qquad \dd \star F_{(0)} = 0 \quad. 
\end{align}
such that the R-R sector can be written as
\begin{equation}
\dd \star F_{(2p)} + H_{(3)} \wedge \star F_{(2p + 2)} = 0
\end{equation}
The equations of motions for the dilaton, instead, read
\begin{equation}
\nabla_M \nabla^M \Phi - \nabla_M \Phi \nabla^M \Phi + \frac{1}{4} \mathcal{R} - \frac{1}{8 \times 3!} |H_{(3)}|^2 = 0 \quad,
\end{equation}
whereas those for the metric are
\begin{align}
e^{- 2 \Phi} \bigl (\mathcal{R}_{MN} + 2 \nabla_M \nabla_N \Phi - & \frac{1}{4} H_{MPQ}{H_N}^{PQ} \bigr ) - \frac{1}{2} (F_{(2)}^2)_{MN} - \frac{1}{2 \times 3!} (F_{(4)}^2)_{MN} + \notag \\  + & \frac{1}{4} G_{MN} \bigl (|F_{0}|^2 + \frac{1}{2!} |F_{(2)}|^2 + \frac{1}{4!}|F_{(4)}|^2 \bigr ) \quad.
\end{align}

\section{Review of the D\texorpdfstring{$\boldsymbol{p}$}{p}-branes effective action (bosonic part)} \label{appendix:brane_action}
As already anticipated in \eqref{eq:DBI}, \eqref{eq:WZ}, the effective action of a stack of D$6$-branes can be decomposed in two pieces
\begin{equation*}
    S^{\textup{DBI}}_{\textrm{D}p}= -T_{\textrm{D}p} \int_{\textup{WV}(\textrm{D}p)} \dd^{p+1} x \; \Tr \bigg( e^{-\hat{\Phi}} \sqrt{-\det(\mathbb{M}_{MN}) \det(\tensor{\mathbb{Q}}{^i_j})} \; \bigg) \quad,
\end{equation*}
\begin{equation*}
    S^{\textup{WZ}}_{\textrm{D}p}= \mu_{\textrm{D}p} \int_{\textup{WV}(\textrm{D}p)}  \Tr \bigg\{ \mathrm{P} \bigg[ e^{i \lambda \iota_Y \iota_Y} \Big(\boldsymbol{\hat{C}} \wedge e^{\hat{B}_{(2)}} \Big) \wedge e^{\lambda \mathcal{F}}  \bigg] \bigg\} \quad.
\end{equation*}
In this appendix we report the key information about these actions, following the notation of \cite{Choi:2018fqw}.
The matrices $\mathbb{M}_{MN}$ and $\tensor{\mathbb{Q}}{^i_j}$ appearing in the DBI action are
\begin{equation} \label{eq:MMN}
    \mathbb{M}_{MN}= \mathrm{P} \Big[ \hat{E}_{MN} + \hat{E}_{Mi} (\mathbb{Q}^{-1} - \delta )^{ij} \hat{E}_{jN} \Big] + \lambda \mathcal{F}_{MN} \quad,
\end{equation}
\begin{equation} \label{eq:Qij}
    \tensor{\mathbb{Q}}{^i_j}=\tensor{\delta}{^i_j}+i \lambda [Y^i,Y^k] \hat{E}_{kj} \quad,
\end{equation}
where
\begin{equation} \label{eq:EMN}
    \hat{E}_{\mathcal{M} \mathcal{N}}=\hat{g}_{\mathcal{M} \mathcal{N}}+\hat{B}_{\mathcal{M} \mathcal{N}} \quad .
\end{equation}
In the expression of the WZ term, $\boldsymbol{C}$ is meant to be a poliform containing the sum of all the $C_{(p)}$ fields, $\iota_{Y}$ stands for the interior product by a vector $Y^i$ and $\mathcal{F}$ is the field-strength of the gauge field $\mathcal{A}$ living on the brane 
\begin{equation} \label{eq:Nonab_FieldStrength}
    \mathcal{F} = \mathrm{d} \mathcal{A} + i \mathcal{A} \wedge \mathcal{A} \;.
\end{equation}
In the above formulas the convention for the 10-dimensional coordinates $x^{\mathcal{M}}$ is that they are split in worldvolume coordinates $x^M$ (which include spacetime coordinates $x^{\mu}$ and internal coordinates $x^a$) and transverse coordinates $y^i$. The position of the non-Abelian $D6$-branes is then encoded in $y^i= \lambda Y^{Ii} t_{I}$, with
\begin{equation} \label{eq:GeneratorRelations}
    [t_I, t_J] = -i {g_{IJ}}^K t_K \quad, \quad \mathrm{Tr} [t_I, t_J] = \delta_{IJ} \quad.
\end{equation}
The symbol $\mathrm{P} [\ldots] $ denotes the pullback of the bulk fields over the worldvolume of the $D6$-branes. As an example, the pullback of $\hat{E}_{MN}$ is
\begin{equation} \label{eq:PullBack}
    \mathrm{P}[\hat{E}_{MN}]= \hat{E}_{MN}+ \lambda D_{M} Y^i \hat{E}_{iN} + \lambda D_{N} Y^i \hat{E}_{Mi} +\lambda^2  D_{M} Y^i D_{N} Y^j \hat{E}_{ij}  \;.
\end{equation}
Here the covariant derivative is defined as
\begin{equation} \label{eq:Covariant_Derivative}
    D_{M} Y^i \equiv \partial_{M} Y^i - i [\mathcal{A}_{M}, Y^i] \;.
\end{equation}
All the hatted fields are meant to be the fields evaluated at the position of the $D6$-branes with a Taylor expansion around $y^i=0$:
\begin{equation}
    \hat{\Phi}(x^M, y^i)= \sum_{k=0}^{\infty} \frac{\lambda^k}{k!} Y^{i_1}\cdots Y^{i_k} \partial_{i_1} \cdots \partial_{i_k} \Phi(x^M, y^i) \Big|_{y^i=0} \;.
\end{equation}

\section{Normalized masses for the solutions in the origin}
\label{appendix_masses}
In this section we describe in detail the results of the computation of the normalized masses for the families $\mathbf{A}$ and $\mathbf{B}$ of solutions found in the origin of the scalar manifold with $\mathfrak{N}=3$. The following table \ref{Masses_A_B} shows the eigenvalues of the normalized mass matrices with the corresponding multiplicities; where the analytic value cannot be obtained explicitly from the Mathematica program, we report its numerical approximation.

\begin{table}[H]
\begin{minipage}{0.47\linewidth}
\centering
\(\begin{array}{cc}
    \toprule
        \multicolumn{2}{c}{\mathbf{A}}\\[3pt]
    \midrule
        m_1^2  \approx -0,999727 & 1 \\[4pt]
        m_2^2 \approx -0,715694 & 5 \\[4pt]
        m_3^2 \approx -0,592392 & 5 \\[4pt]
        0 & 13 \\[4pt]
        \frac{5}{2} - \frac{\sqrt{65}}{6} & 3 \\[4pt]
        m_4^2 \approx 1,44514 & 5 \\[4pt]
        m_5^2 \approx 1,86103 & 5 \\[4pt]
        2 & 1 \\[4pt]
        \frac{1}{24} \big( 77 -3 \sqrt{65}\big) & 3 \\[4pt]
        \frac{1}{24} \big( 65 + \sqrt{65}\big) & 1 \\[4pt]
        \frac{1}{24} \big( 41 + 9 \sqrt{65}\big) & 1 \\[4pt]
        m_6^2 \approx 5,29015 & 1 \\[4pt]
        m_7^2 \approx 5,66038 & 5 \\[4pt]
        \frac{20}{3} & 1 \\[4pt]
        m_8^2 \approx 6,81315 & 5 \\[4pt]
        m_9^2 \approx 7,59755 & 1 \\[4pt]
    \bottomrule
\end{array}\)   
\end{minipage}
\begin{minipage}{0.47\linewidth}
\centering
\(\begin{array}{cc}
    \toprule
        \multicolumn{2}{c}{\mathbf{B}}\\[3pt]
    \midrule
        \frac{1}{24} \big( 41 - 9 \sqrt{65}\big) & 1 \\[4pt]
        m_{10}^2  \approx -0,688753 & 5 \\[4pt]
        m_{11}^2 \approx -0,481668 & 1 \\[4pt]
        0 & 13 \\[4pt]
        2 & 1 \\[4pt]
        m_{12}^2 \approx 2,15924 & 5 \\[4pt]
        \frac{1}{24} \big( 65 - \sqrt{65}\big) & 1 \\[4pt]
        \frac{1}{6} \big( 15 +\sqrt{65}\big) & 3 \\[4pt]
        m_{13}^2 \approx 3,8825 & 5 \\[4pt]
        \frac{1}{24} \big( 77 +3 \sqrt{65}\big) & 3 \\[4pt]
        m_{14}^2 \approx 4,24021 & 1 \\[4pt]
        m_{15}^2 \approx 4,77015 & 1 \\[4pt]
        m_{16}^2 \approx 5,35396 & 5 \\[4pt]
        \frac{20}{3} & 1 \\[4pt]
        m_{17}^2 \approx 9,23833 & 5 \\[4pt]
        m_{18}^2 \approx 12,6664 & 5 \\[4pt]
    \bottomrule
\end{array}\)   
\end{minipage}

\caption{\emph{Normalized masses for the families $\mathbf{A}$ and $\mathbf{B}$ of solutions, for any choice of the signs $s_1$ and $s_2$.}}
\label{Masses_A_B}
\end{table}

The values that are only provided numerically in table \ref{Masses_A_B} can be however expressed as the roots of three polynomials:
\begin{equation}
\label{Polynomials_massesAB}
    \begin{split}
       P_1= & \, 4566016 + 10548720 \, x - 287910 \, x^2 - 4337091 \, x^3 + 1686420 \, x^4 - 
 238140 \, x^5 \\
        &+ 11664 \, x^6 \;, \\
        P_2 = &-11239424 + 972288 \, x + 14286096 \, x^2 - 10426914 \, x^3 + 2802519 \, x^4 - 
 311040 \, x^5  \\
        & +11664 \, x^6 \;, \\
        P_3 = & \, 7702016 + 15221568 \, x - 6194124 \, x^2 - 12736413 \, x^3 + 6248988 \, x^4 - 952560 \, x^5  \\
        &+46656 \, x^6 \;. \\
    \end{split}
\end{equation}

In table \ref{Roots_massesAB} we specify, for any of the masses that are given numerically in table \ref{Masses_A_B}, which polynomial among the three in \eqref{Polynomials_massesAB} it is a root of, showing the roots in increasing order.

\begin{table}[H]
\begin{minipage}{0.3\linewidth}
\centering
\(\begin{array}{cc}
    \toprule
        \multicolumn{2}{c}{P_1}\\[3pt]
    \midrule
        m_1^2  & 1^{\textup{st}} \text{  root} \\[4pt]
        m_{11}^2  & 2^{\textup{nd}} \text{  root} \\[4pt]
        m_{14}^2  & 3^{\textup{rd}} \text{  root} \\[4pt]
        m_{15}^2  & 4^{\textup{th}} \text{  root} \\[4pt]
        m_6^2  & 5^{\textup{th}} \text{  root} \\[4pt]
        m_9^2  & 6^{\textup{th}} \text{  root} \\[4pt]
    \bottomrule
\end{array}\)   
\end{minipage}
\begin{minipage}{0.3\linewidth}
\centering
\(\begin{array}{cc}
    \toprule
        \multicolumn{2}{c}{P_2}\\[3pt]
    \midrule
        m_2^2  & 1^{\textup{st}} \text{  root} \\[4pt]
        m_5^2  & 2^{\textup{nd}} \text{  root} \\[4pt]
        m_{12}^2  & 3^{\textup{rd}} \text{  root} \\[4pt]
        m_{13}^2  & 4^{\textup{th}} \text{  root} \\[4pt]
        m_8^2  & 5^{\textup{th}} \text{  root} \\[4pt]
        m_{18}^2  & 6^{\textup{th}} \text{  root} \\[4pt]
    \bottomrule
\end{array}\)   
\end{minipage}
\begin{minipage}{0.3\linewidth}
\centering
\(\begin{array}{cc}
    \toprule
        \multicolumn{2}{c}{P_3}\\[3pt]
    \midrule
        m_{10}^2  & 1^{\textup{st}} \text{  root} \\[4pt]
        m_3^2  & 2^{\textup{nd}} \text{  root} \\[4pt]
        m_4^2  & 3^{\textup{rd}} \text{  root} \\[4pt]
        m_{16}^2  & 4^{\textup{th}} \text{  root} \\[4pt]
        m_7^2  & 5^{\textup{th}} \text{  root} \\[4pt]
        m_{17}^2  & 6^{\textup{th}} \text{  root} \\[4pt]
    \bottomrule
\end{array}\)   
\end{minipage}

\caption{\emph{Correspondence between normalized masses of the families $\mathbf{A}$ and $\mathbf{B}$ of solutions and the roots of the polynomials in \eqref{Polynomials_massesAB}.}}
\label{Roots_massesAB}
\end{table}

\bibliographystyle{jhep}
\bibliography{bibliography.bib}

\end{document}